\documentclass[
twocolumn,
reprint,
 amsmath,amssymb,
 aps,
 prc,
]{revtex4-2}

\usepackage{natbib}
\usepackage{bm}
\usepackage{dcolumn}
\usepackage{graphicx}
\usepackage[utf8]{inputenc}
\usepackage[english]{babel}

\newenvironment{equationc}{\begin{equation}}{,\end{equation}\ignorespacesafterend}
\newenvironment{equationp}{\begin{equation}}{.\end{equation}\ignorespacesafterend}

\begin{document}


\title{AI-Assisted analysis of $^{28}$Si$^*$ $\rightarrow$ 7$\alpha$ break-up data}


\author{ T. Depastas$^{a,*}$, A. Bonasera$^{a,b}$ and J. Natowitz$^{a}$}


\affiliation{ $^{a}$ Cyclotron Institute, Texas A\&M University,
                     College Station, Texas, USA }
\affiliation{ $^{b}$ Laboratori Nazionali del Sud, INFN, Catania 95123, Italy }

\address{ $^{*}$ Corresponding author. Email:  tdepastas@tamu.edu}

\begin{abstract}
Mid-weight $\alpha$-conjugate nuclei are predicted to possess exotic toroid-like resonances with high angular momenta. The search for these states in $^{28}$Si$^*$ is the main point of two published experimental investigations of the peripheral $^{28}$Si + $^{12}$C reaction by Cao and collaborators and by Hannaman and collaborators. In this work, we develop a novel Artificial Intelligence (AI)-based machine learning method utilizing the Gaussian Mixture Model (GMM) to analyze available experimental and theoretical data. We additionally study the reaction with the Hybrid $\alpha$-Cluster (H$\alpha$C) model. In all the examined data our results suggest the presence of underlying structure which is close to that predicted for toroidal states.
\end{abstract}


\maketitle

\section{Introduction}
\label{s1}
Due to their high binding energy \cite{NNDC2022}, alpha particles, tend to appear as clusters in nuclei \cite{Ikeda1968}. This effect can be profound in a wide range of phenomena, ranging from low-energy astrophysical reactions, structure-as in the Hoyle state of carbon-12 \cite{Hoyle1954}, descriptions of super-novae \cite{Borderie2021} and rotational bands of deformed nuclei \cite{Ikeda1968}, to the equation of state of $\alpha$-clustered Nuclear Matter \cite{Brink1973,BonaseraAlphaMatter2011}.\par
The presence of toroid-like excited states was first predicted by Bohr, Mottelson, Wheeler, Nilsson, Rainwater and later Wong \cite{Bohr1951,BohrMottelson1953,Wheeler1957,Rainwater1950,Wong1972}. More specifically, Wong studied these structures in light to mid-weight nuclei with high angular momenta from the Liquid Drop \cite{Wong1973,Wong1978} and Stasczak and Wong predicted the existence of toroidal states in many light nuclei \cite{Wong2014}.\par
Searches of toroidal states in silicon-28 have been carried out in two recent experiments in the Texas A\& M Cyclotron Institute, experimentally facilitated using the inverse kinematics reaction $^{28}$Si + $^{12}$C at $35$ MeV/A beam energy. The existence of toroidal states with high angular momenta and a subsequent dissociation into $7\alpha$ particles was explored by a search for resonant peaks in the
excitation energy spectrum, i.e., the differential $7\alpha$ cross section as a function of the excitation energy projectile-like $^{28}$Si fragments produced in mid-peripheral collisions. The first experiment, by Cao and collaborators used the $4\pi$ Neutron Ion Multidetector for Reaction Oriented Dynamics (NIMROD) combined with the Indiana Silicon Sphere (ISiS) and reported the presence of possible toroid resonances at 114, 126 and 138 MeV \cite{Cao7aPRC}. These results were supported by theoretical Shell Model and Relativistic Density Functional Theory (DFT) calculations
\cite{Cao7aPRC,Ren2020}. The second experiment, by Hannaman and collaborators utilized the Forward Array Using Silicon Technology (FAUST) employing dual-axis duolateral (DADL) detectors \cite{Hannaman7aPRC,HannamanEPJ}. They concluded that there no statistically significant peaks in the $7\alpha$ excitation spectrum above the ``background". The kinematic acceptances of the two different detection systems lead to somewhat different excitation energy spectra. Two subsequent works \cite{WadaCimento2025,NatowitzTalk} have argued that there are, in fact, resonant peaks in the Hannaman et al. data. \par
To probe this problem theoretically, the Hybrid $\alpha$-Cluster (H$\alpha$C) model was introduced in Ref. \cite{ZhengHac2021} and later used for sub-barrier fusion studies \cite{Depastas2023,Depastas2024EPJ,Depastas2024Plb}. In the model initial quantized angular momentum was given to the $^{28}$Si. This resulted in structures in the final cross section, which correspond to toroidal states, whose excitation energies depend critically on the interaction \cite{ZhengHac2021,Bass1977}.\par
With this work, we aim to further contribute to the discussion by analyzing the data with a novel Artificial Intelligence (AI)-based method, while also noting the drawbacks of the previously used polynomial fit method \cite{HannamanEPJ}. AI-inspired methods, ranging from Support Vector Machines \cite{DeSanctis2009} to Machine Learning algorithms have taken traction in the past two decades, with the latter being the basis of the work awarded the 2024 Nobel Prizes for both Physics \cite{Hopfiled1982Nobel2024,Ackley1985Nobel2024} and Chemistry \cite{Hassabis2021Nobel2024}. In this work we first “train” the AI algorithm by studying the rotating silicon H$\alpha$C data of Ref. \cite{ZhengHac2021}. We then, apply the method to the experimental data from the literature, as well as to results of new theoretical H$\alpha$C calculations for the reacting system. The structure of the paper is as follows. In section \ref{s3}, we discuss the novel machine learning method and compare our results to the previous analyses. In section \ref{s2} we describe the details of the new theoretical calculations and present the resulting excitation spectra. The results of our analysis are shown in section \ref{s4}, while in section \ref{s5} we present our conclusions.
\section{Polynomial and Machine Learning Methods}
\label{s3}
The search for toroidal resonances in Ref.s \cite{Cao7aPRC} and \cite{Hannaman7aPRC} proceeds via the subtraction of a continuum background from the obtained excitation energy spectrum. In the former, this background is taken as a random average of mixed events and in the latter, as a $9^{th}$ order polynomial fit. The peaks are then identified based on the statistically significant differences between the background and the data. The polynomial method of Ref. \cite{Hannaman7aPRC} has a couple important drawbacks. First, any peak can be fit with a non-linear polynomial, given enough terms. This in turn leads to a loss of possible real peaks in the subtraction. Second, a polynomial function possesses several ``special" points of maxima, minima and inflections that are given by the roots of each derivative of the polynomial. Since a polynomial of order $N$, can in principle have at most $N$ roots, a subtractive analysis yields additional structure that is not present in the data. In our analysis, we obtain the ``special" points numerically for each dataset, by fitting a $9^{th}$ order polynomial and accept the roots on each derivative if they are real and their value of the next derivative is negative ($2^{nd}$ Derivative Criterion).\par
\begin{figure}[ht]                                        
\includegraphics[width=8 cm]{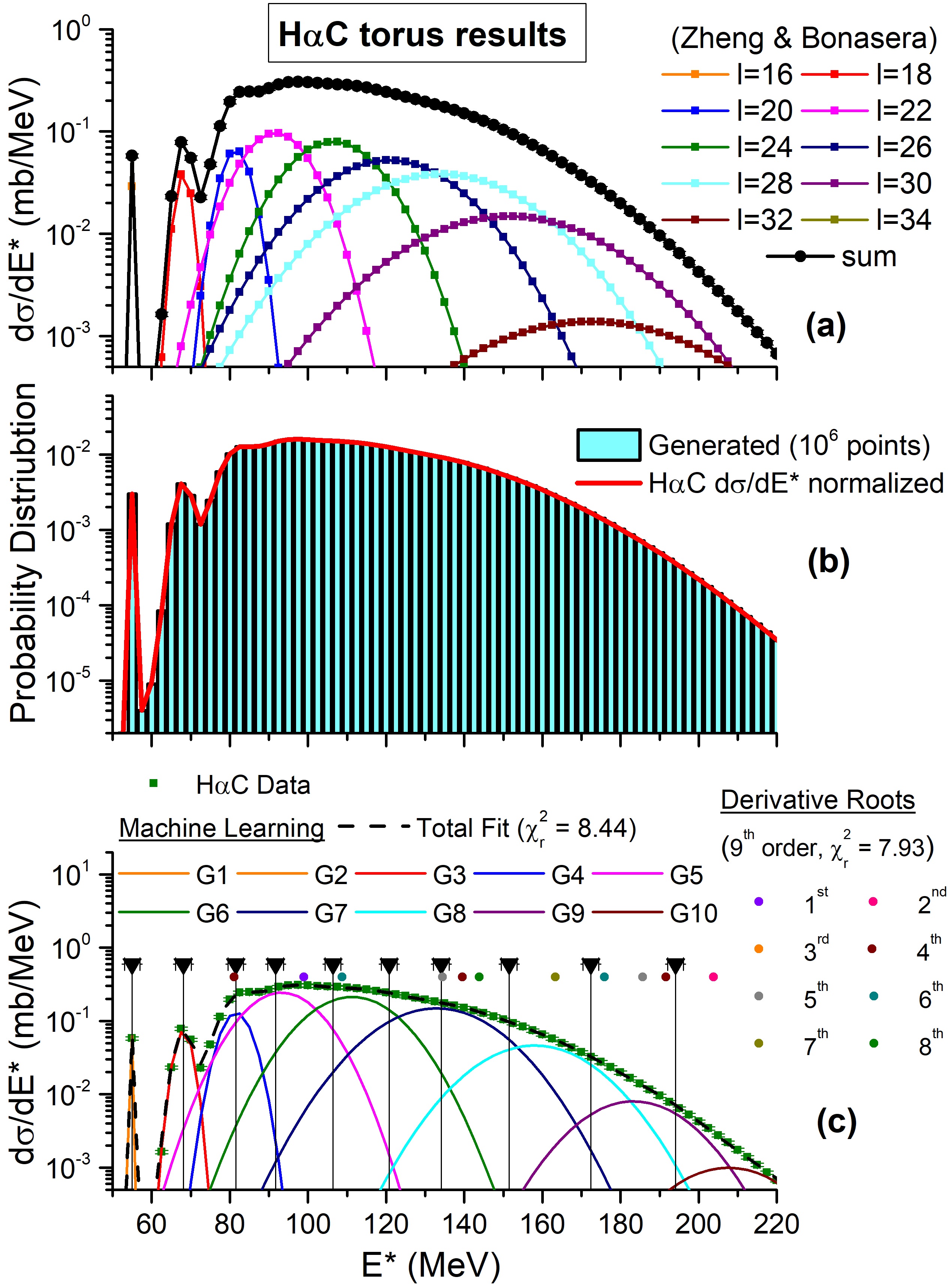}
\caption{(Color online) Panel (a): Original H$\alpha$C toroid data and their Gaussian components, according to the key (reproduction of Fig. 11 of Ref. \cite{ZhengHac2021}). The $l=34$ peak is below the scale. Panel (b): Comparison of generated distribution of $10^6$ points with the H$\alpha$C cross section compared to unity. Panel (c): Machine Learning Gaussians as full lines (termed $Gi$, $i=1,...,10$), their linear combination as dashed line and Polynomial Maxima roots as full dots (order of derivative according to color). The square points represent the H$\alpha$C toroid data and the triangles show the centroids of their Gaussian constituents.}
\label{Fig3}
\end{figure}

The analytical method we propose here is substantially different and is inspired by the approach of Ref. \cite{ZhengHac2021}. There, the authors calculate the average and standard deviation of the excitation energy of a rotating silicon for each quantized angular momentum value, with the H$\alpha$C model. The differential cross section, is then, written as a sum of Gaussian components (along with an exponential phase-space factor) from the extracted means and standard deviations. Their results are reproduced in panel (a) of Fig. \ref{Fig3}. Here, we attempt to apply a similar process to data with unknown underlying structure. We consider a general graph with equally spaced points with error-bars $\left\{x_i,y_i,\delta y_i \right\}$. Our goal is to break down this graph into a linear combination of $N_G$ Gaussian functions, i.e.,
\begin{equationc}
y_i\pm\delta y_i \approx \sum_{I_G=0}^{N_G-1}w_{I_G}g_{I_G}\left(x_i;\mu_{I_G},\sigma_{I_G}\right) \text{ } \forall i
\label{e9}
\end{equationc}
where each Gaussian $g_{I_G}$ is normalized to unity and is characterized by its mean $\mu_{I_G}$, its width $\sigma_{I_G}$ and a mixing coefficient $w_{I_G}$. The values of these variables are determined by employing a modified version of the Unsupervised Machine Learning technique Gaussian Mixture Model (GMM) \cite{GMMBOOK}, utilizing the {\it Scikit-learn} Python library \cite{scikit-learn}. Firstly, the points are binned with an equal spacing along their x-values, if they are not already equally spaced and then, are transformed to a probability distribution by normalizing to unity. From this discrete histogram, we generate $N_p=10^6$ random points, from a normal distribution around the center of each bin. An example of such a generated distribution is shown in Fig. \ref{Fig3}, panel (b) for the data of Ref. \cite{ZhengHac2021}. The problem at this point is to map each point to one of the Gaussians and in turn determine their characteristic variables. For this, we use the library's built-in Expectation-Minimization (EM) algorithm, which iteratively groups the points into Gaussians and calculates their characteristics until convergence \cite{GMMBOOK}. The initial values for $\left\{\sigma_{I_G}\right\}$ and $\left\{w_{I_G}\right\}$ are randomly chosen by the library's ``k++ means" method \cite{kmeans}, while for the means, we adopt the following scheme for $I_G=0,...,N_G-1$:
\begin{equationp}
   \mu_{I_G}^{(0)}= \left(x_{max}-x_{min}\right)\frac{I_G\left(I_G+1\right)}{N_G\left(N_G-1\right)}+x_{min}
\label{e10}
\end{equationp}
This ensures that the initial mean values belong in the range $x_{min}=\underset{i}{\min}\left(x_i\right)$ and $x_{max}=\underset{i}{\max}\left(x_i\right)$ of the data. The optimal number of Gaussians $N_G$ to use is found by restricting the value of Reduced Chi-Square $\chi^2_{\nu}$ of the fit to be close to 1 \cite{chi2Ref}. The statistics are defined as:
\begin{equationp}
\chi^2_{\nu}=\frac{\sum_i\left\{\frac{1}{\delta y_i}\left[y_i - \sum_{I_G=0}^{N_G-1}w_{I_G}g_{I_G}\left(x_i;\mu_{I_G},\sigma_{I_G}\right)\right]\right\}^2}{N_p-3 N_G}
\label{e11}
\end{equationp}
\par
We follow the method described above with the excitation energies as the x-values and the differential cross sections and/or counts per MeV (yield) as the y-values. The resulting Gaussian peaks correspond to components in the excitation spectrum. We stress that here, we do not use the Machine Learning method in a predictive way, but as an analytical tool to extract structure which may be already present in the data.\par
Since the H$\alpha$C \cite{ZhengHac2021} are composed of known Gaussians ( reproduced in Fig. \ref{Fig3} panel a), we use them to test our method. We present the machine learning and polynomial results in Fig. \ref{Fig3} (panel c). The square points in green correspond to the original H$\alpha$C data and the ten black triangular points correspond to the centroids of their Gaussian components. The full lines represent the ten Gaussians resulting from the machine learning process (according to the key) and the dashed line their linear combination. The full dots signify the maxima roots of the polynomial derivatives. For the calculation of the $\chi^2_{\nu}$ we assume a uniform $5\%$ error, which results into similar values for both polynomial and machine learning methods.\par
We observe that seven out of ten machine learning Gaussians reproduce the known peaks compared to four out of twelve polynomial roots. This demonstrates the superiority of the proposed AI method. We are also able to extract close peaks in the low $E^*$ region, where the cross section is smooth and a polynomial fitting method fails. The accuracy limits of our method are challenged in the regions of low statistics and for peaks with small mixing coefficients. These situations are of course problematic for all fitting methods.
\section{\MakeLowercase{c}H$\alpha$C Reaction Calculations}
\label{s2}
In addition to the proposed AI-based analytical method, we utilize the cH$\alpha$C model (that is the H$\alpha$C model for collisions) to expand the available theoretical results on the $^{28}$Si + $^{12}$C reaction. The model \cite{ZhengHac2021} adopts a molecular dynamics approach with $\alpha$ clusters as fundamental degrees of freedom. These interact via the Coulomb and Bass \cite{Bass1977} potentials, while the quantum Heisenberg and inner Pauli correlations are approximated by an effective repulsive Fermi energy \cite{ZhengHac2021}.\par
First, the ground state of each nucleus is calculated and used as the initial condition for the Hamiltonian equations of motion. Then, the $\alpha$ positions are randomly rotated to produce a random event and the nuclei collide with 35 MeV/A beam energy in inverse kinematics (294 MeV in the C.M. frame). The impact parameters of each event are quantized with an initial $l$-value according to the relation,
\begin{equationc}
    l\hbar=bp_{C.M.}=b\sqrt{2\mu E_{C.M.}}
\label{e1}
\end{equationc}
on the reaction ($z$-)axis. Since the nuclei are non-identical, $l$ can take any non-negative integer value. The system evolves for a $800$ fm/c and the resulting fragments are collected. The identification of fragments follows a simple algorithm similar to the one used in the Constrained Molecular Dynamics (CoMD) model \cite{BonaseraCoMD}, where a pair of particles at most $6$ fm apart belong to the same fragment. Here, we combine fragments from $40400000$ events, with $0 \le l \le 50$. We find that the maximum initial angular momentum for which $7\alpha$ particles are produced is $48\hbar$. We stress that the goal of these calculations is not to describe the physical information in detail, but to provide a test ground to train our AI approach.  In any case, as we show subsequently, the model describes part of the data rather well.\par
The next step is the choice of the $7\alpha$ events. As discussed in Ref. \cite{ZhengHac2021}, the $^8$Be is slightly overbound in the H$\alpha$C model, in order to accurately parametrize the ground state energies of a wide range of nuclei. That leads us to account the beryllium-8 nuclei along with the ``free" $\alpha$ particles towards the fragments of the candidate events. Furthermore, in many cases, the carbon nucleus also breaks into its $\alpha$ constituents, resulting in events with more than $7$ $\alpha$ particles. To choose the $7\alpha$ fragments (``free" or bound in a $^8$Be) that contribute to the cross section we follow two different approaches. One approach is to consider only the $\alpha$ particles that originally were belonging to the silicon-28 projectile (Filter 1) and the other is to collect the fragments in the forward direction in the C.M. frame, i.e., with momentum $p_z > 0$ in the model beam $z$-axis (Filter 2). The $^{28}$Si excitation energy of each $7\alpha$ event is then, calculated via the formula:
\begin{equationc}
    E^*=E_{k,\alpha}+E_{k,^{8}Be}+E_{C}-E_{C.M.,Si}-E_{7\alpha,Si}
\label{e2}
\end{equationc}
where $E_{k,\alpha}$ and $E_{k,^{8}Be}$ are the kinetic energies of the ``free" and $^8$Be $\alpha$'s in the C.M. of the total system, $E_{C.M.,Si}$ is the C.M. kinetic energy of the excited $^{28}$Si, $E_{7\alpha,Si}$ is H$\alpha$C calculated Q-value for the $7\alpha$ break-up and $E_{C}$ is the electrostatic energy between the fragments, which is found to be less than $1$ MeV in total. The term $E_{C.M.,Si}$ is subtracted such that the excitation energy is calculated in the frame of $^{28}$Si$^*$, instead of the total $^{28}$Si + $^{12}$C system. The terms of Eq. \ref{e2} are defined as:
\begin{equationc}
    E_{k,\alpha}=\frac{1}{2 m_{\alpha}}\sum_{i \in \alpha}p_i^2
\label{e3}
\end{equationc}
\begin{equationc}
    E_{k,^{8}Be}=\frac{1}{2 m_{\alpha}}\sum_{i \in ^{8}Be}\lambda_i\left[\left(\frac{\bm{p}_i}{2}\right)^2+\frac{E^*_{^{8}Be,i}}{2}\right]
\label{e4}
\end{equationc}
\begin{equationc}
    E_{C}=\sum_{i<j \in \alpha, ^{8}Be}\frac{Z_iZ_je^2}{r_{ij}}
\label{e5}
\end{equationc}
\begin{equationc}
    E_{C.M.,Si}=\frac{1}{14 m_{\alpha}}\left(\sum_{i \in \alpha}\bm{p}_i+\sum_{i \in ^{8}Be}\frac{\lambda_i}{2}\bm{p}_i\right)^2
\label{e6}
\end{equationc}
\begin{equationp}
    E_{7\alpha,Si}=E_{Si}-7 E_{\alpha}
\label{e7}
\end{equationp}
\par
In the previous equations, $\bm{p}_i$ refers to the C.M. momenta of the $\alpha$'s and $^{8}$Be's, $Z_{i/j}=2,4$ the atomic numbers and $r_{ij}$ the relative distances of the fragments. In Eq.s \ref{e4} and \ref{e6}, the factor $\lambda_i=0,1,2$ is the number of $\alpha$'s to be taken into account in the beryllium nucleus, according to the aforementioned two filters. With Filter 1, $\lambda_i$ is the number of $\alpha$'s that originate from the projectile and with Filter 2, $\lambda_i=2$ if the C.M. momentum of beryllium is positive and $\lambda_i=0$, otherwise. We assume that the C.M. momentum and excitation energy $\left( E^*_{^{8}Be,i}\right)$ of a decaying beryllium-8 is distributed equally to its $\alpha$ constituents. The Q-value term of Eq. \ref{e7} is calculated via the ground state energies of the silicon-28 and helium-4 nuclei given by the initial conditions of the H$\alpha$C calculation and is within the $2\%$ error of the model \cite{ZhengHac2021,Depastas2023}.\par
The events with a given $l$-value and excitation energy $E^*$ are accordingly binned, with $\delta E^*=2$ MeV bin width and the differential cross section $\frac{d\sigma}{d E^*}$ with error $\frac{\delta\sigma}{d E^*}$ is calculated by the formula:
\begin{equationc}
    \frac{d\sigma\left(\pm\delta\sigma\right)}{d E^*}\approx\frac{\pi\hbar^2}{2\mu E_{C.M.}}\sum_{l}\left(2 l+1\right)\frac{\delta N_l \left( \pm \sqrt{\delta N_l} \right) }{N_l \delta E^*}
\label{e8}
\end{equationc}
where $N_l$ is the total number of calculated events with specified $l$-value, from which the number $\delta N_l$ of $7\alpha$ events is selected for each bin, with the standard Poisson $\sqrt{\delta N_l}$ error.\par
\begin{figure*}[ht]                                        
\includegraphics[height=10.0 cm]{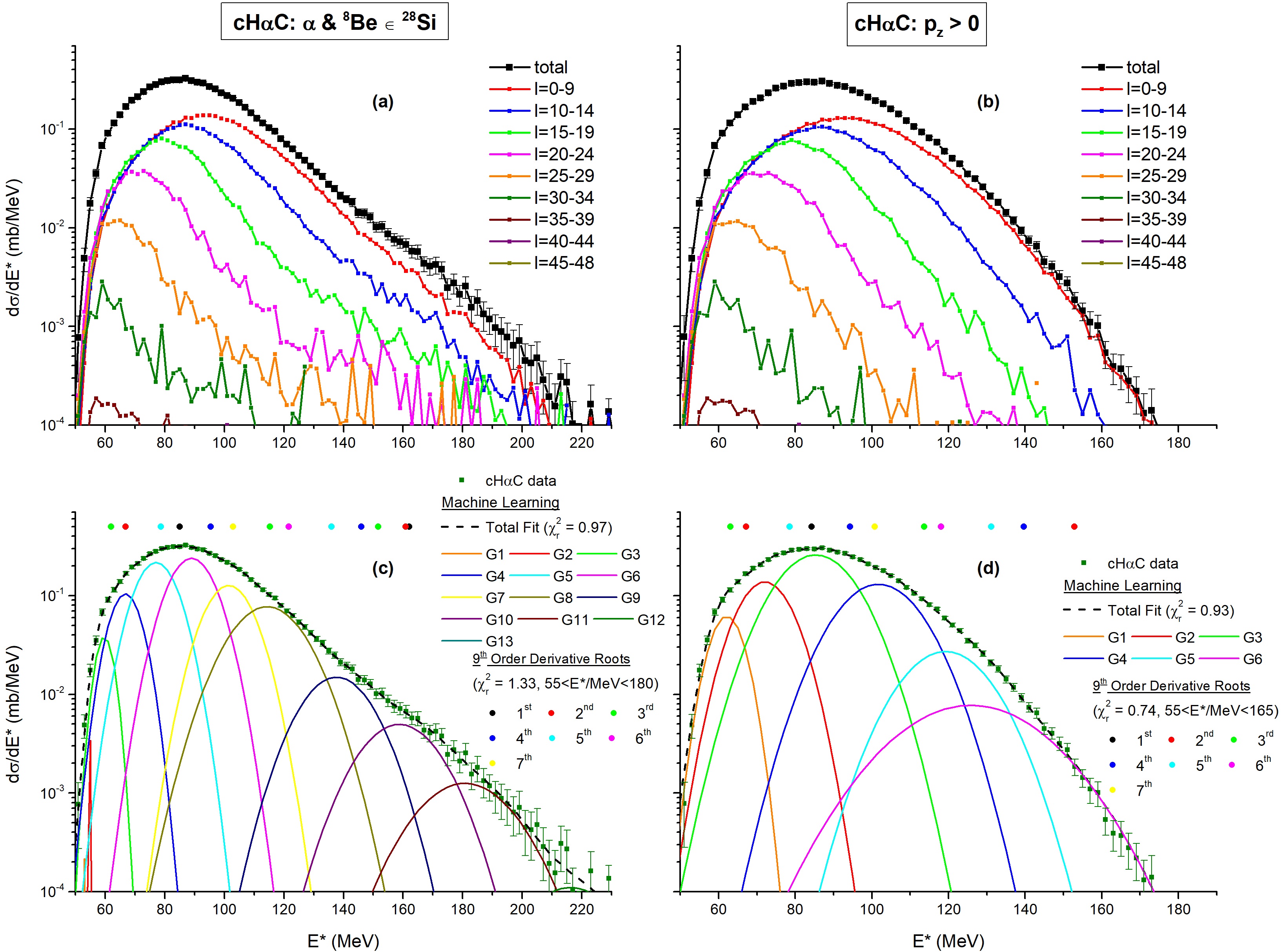}
\caption{(Color online) Partial and total differential cross sections for different ranges of $l$-values (panels a and b). The Machine Learning Gaussian results, their linear combination and the Polynomial Maxima roots, according to the key (panels c and d). The left column corresponds to the $7\alpha$ fragments belonging to Si (Filter 1) and the right column to fragments in the forward direction in the C.M. frame (Filter 2).}
\label{Fig1}
\end{figure*}
In Fig. \ref{Fig1} we present the results of the collision H$\alpha$C calculations and the machine learning analysis. In the left column we show the results where the $7\alpha$ fragments come from the Si (Filter 1) and on the right the fragments in the forward direction in the C.M. frame (Filter 2). In panels (a) and (b) the total (black line) and partial differential cross sections, for different ranges of $l$-values in the sum of Eq. \ref{e8} are plotted, according to the key. We group the partial cross sections for $l=0-9$ and after that with steps of five. The reason for this grouping is that the maximum impact parameter where the projectile and target nuclei fully overlap is $l_{fo}=b_{fo}p_{C.M.}/\hbar=(R_{Si}-R_C)p_{C.M.}/\hbar\approx 9$, where $R_{Si}$ and $R_C$ are the projectile and target radii. Essentially, the curves correspond to the excitation of the overlapping $^{28}$Si$^*$ continuum states. The higher initial partial waves associated with higher impact parameters reduce the interaction area of the reactant system and thus, less excitation energy is transferred \cite{BonaseraDissipative1990}.\par
In panels (c) and (d), we apply the AI and polynomial methods to the model collision data and present the results, for Filters 1 and 2 respectively. The symbols follow the nomenclature of Fig \ref{Fig3}, panel (c). Both filters give quite similar results, except close to $E^*\sim 50$ MeV and above $E^*>150$ MeV, where statistics are low. For that reason, data from Filter 1, yield three more peaks than Filter 2, which are included in wider Gaussians. Nonetheless, both results agree in the toroid excitation energy region of interest $E^*\sim 60-150$ MeV. The shapes of the $l$-value contributions (panels a and b) are different than the machine learning results (panels c and d), since the latter should correspond to the transferred angular momenta to the projectile-like fragment and the former to the initial angular momenta of the reactant system. We note that again, the polynomial fit yield several maxima roots, which cannot be mapped to specific Gaussians. \par
It is interesting to closely observe the H$\alpha$C cross sections, their original constituents and machine learning Gaussian components for the rotating silicon (Fig. \ref{Fig3}) and collision (Fig. \ref{Fig1}) calculations. When the two nuclei collide $\alpha$-$\alpha$ interactions occur and excitation energy is shared between the projectile and target like fragments as thermal and rotational motion. This widens the cross section contributions of the different initial $l$-values and the rotational peaks of silicon, which correspond to machine learning Gaussians. Specifically, higher excitation energies, which come from smaller impact parameters, are characterized by more chaotic motion and result in wider peaks, with the $l=0$ being the widest as it is purely thermal. On the contrary, in Ref. \cite{ZhengHac2021}, the motion is initially purely rotational.\par
\begin{figure}[ht]                                        
\includegraphics[width=8.5cm]{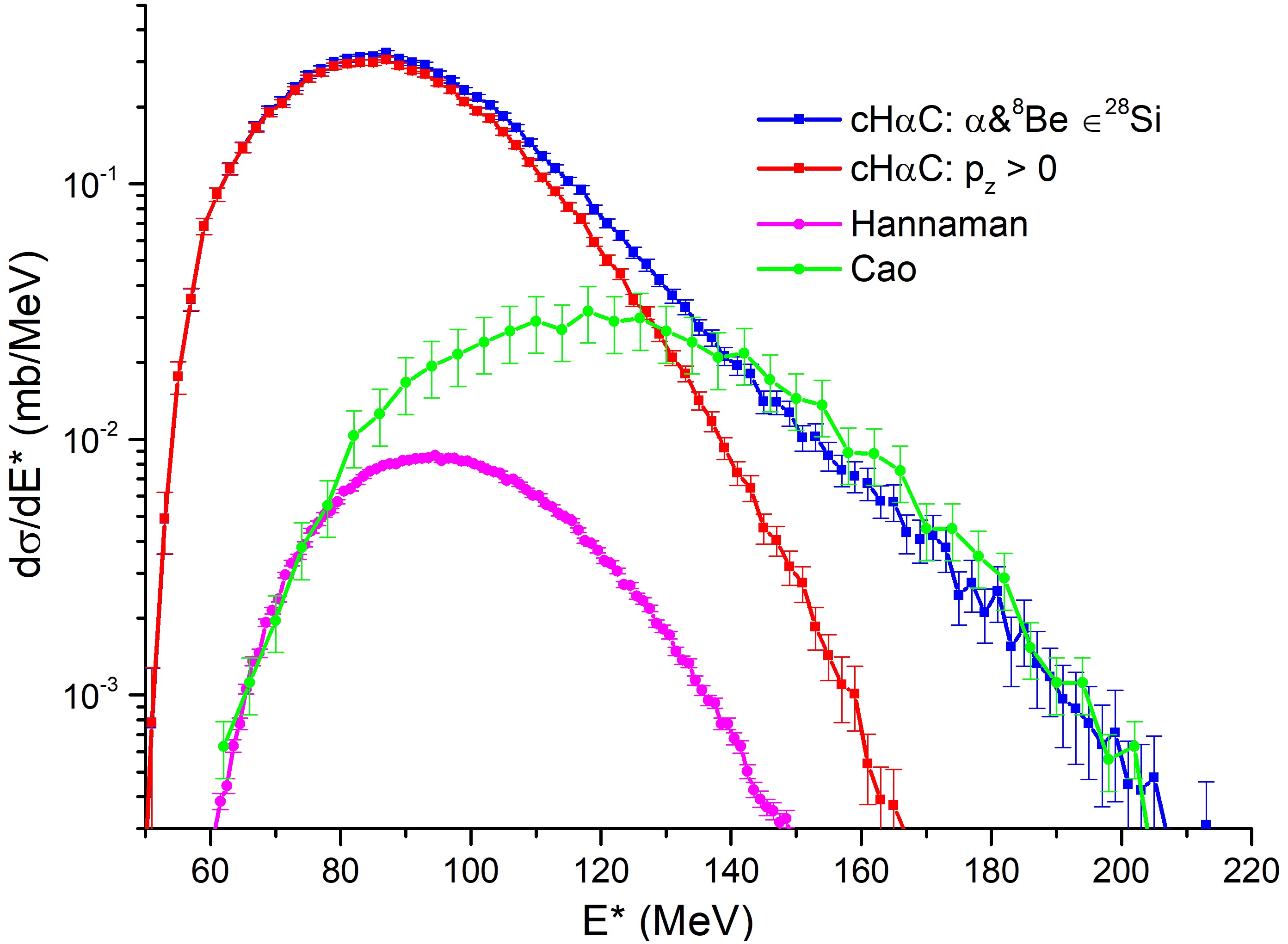}
\caption{(Color online) Differential Cross Sections calculated via the H$\alpha$C model for the reacting system under different filters (blue and red) and experimental datasets. The Cao data (green) are taken from Ref. \cite{Cao7aPRC} and the Hannaman data (magenta) taken from Ref. \cite{Hannaman7aPRC} and scaled by a factor of $2$ $10^{-6}$ (a.u.) to the low energy tail of the Cao data.}
\label{Fig2}
\end{figure}
We compare cH$\alpha$C results with the available experimental data in Fig. \ref{Fig2}. The blue and red lines correspond to the new cH$\alpha$C results utilizing the Filters 1 and 2, respectively, while the toroid $^{28}$Si calculation from Ref. \cite{ZhengHac2021} corresponds to the green curve. The Cao experimental data obtained by Ref. \cite{Cao7aPRC} are shown in green and the Hannaman data in magenta. The latter are taken from Fig. 2 of Ref. \cite{Hannaman7aPRC}, where they are binned with a width of 1.25 MeV and include standard Poisson errors. The transformation from counts to differential cross section occurs via the scaling with a factor of $2$ $10^{-6}$ (mb/MeV) to the low energy region of Cao data \cite{Cao7aPRC}. The acceptance of the two experimental setups is different, which explains their different shapes above $120$ MeV excitation energy.\par
We observe that the new cH$\alpha$C calculation results converge to the experimental data in the high $E^*$ tail, while they are generally higher at lower $E^*$ values. This is consistent with the trend of the toroid silicon calculations \cite{ZhengHac2021}. This implies a high degree of clusterization for the most energetic fragments in the experimental data. On the contrary, for lower excitation energies, there are many open non-$\alpha$-conjugate exit channels not available in our model. In that sense, the theoretical calculations provide an upper limit to the cross sections obtained experimentally. We additionally, note the similarity to the recent experimental data of Ref. \cite{Borderie2021} and the AMD results of Ref. \cite{Cao7aPRC}.\par
Our results possess a vast wealth of interesting physical characteristics, ranging from collective oscillations to hydrodynamical effects. These are beyond the scope of the present work, which we emphasize is the analysis of the available cross sectional data and consequently, will be discussed in a future paper. 
\section{Analysis of Experimental Data}
\label{s4}
\begin{figure*}[ht]                                        
\includegraphics[width=12.5 cm]{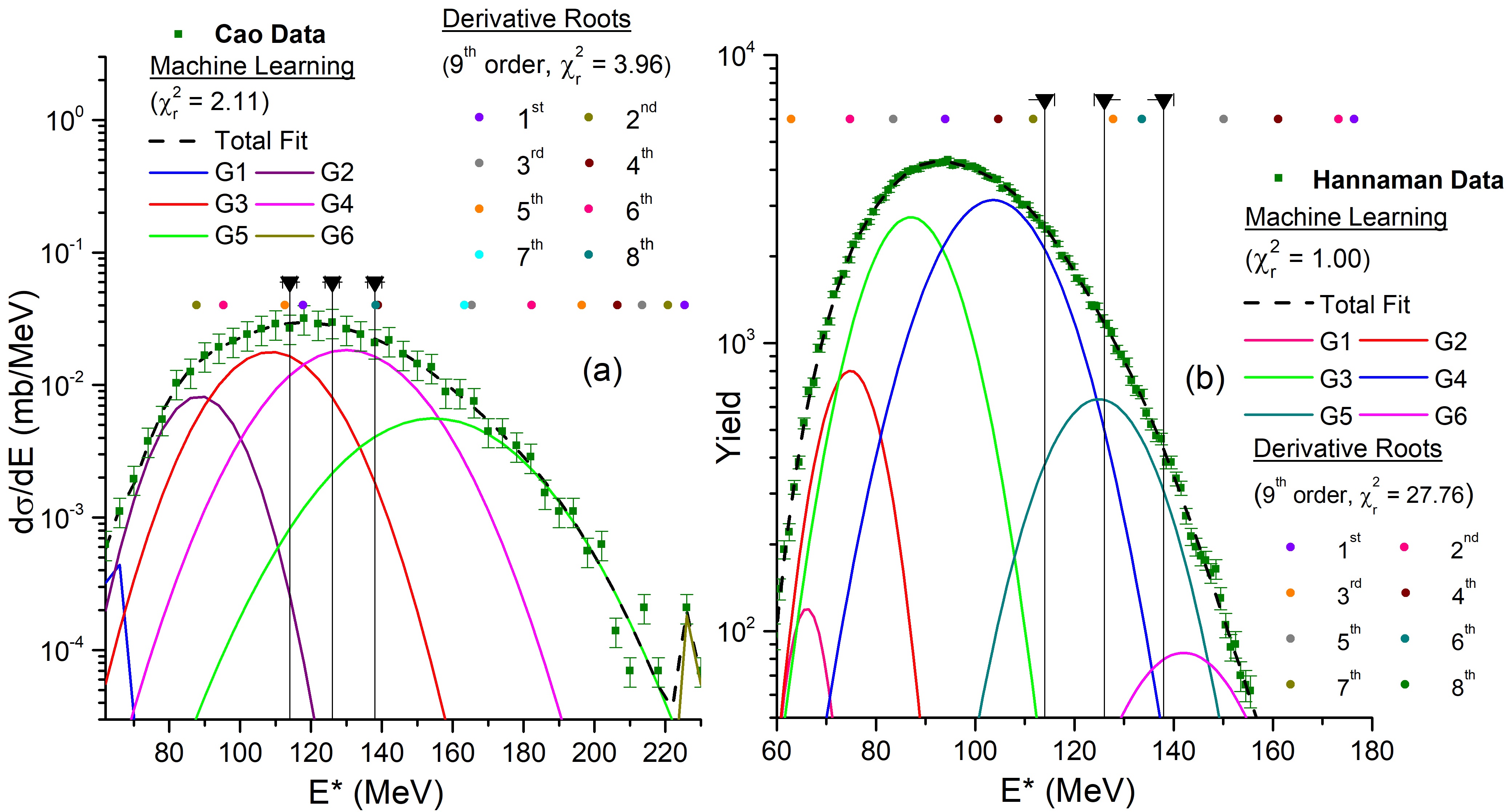}
\caption{(Color online) Panel (a): Results for the Cao \cite{Cao7aPRC} and panel (b) for the Hannaman \cite{Hannaman7aPRC} experimental data. The Machine Learning Gaussians are represented as full lines (termed $Gi$, $i=1,...$), their linear combination as dashed line, the Polynomial Maxima roots as full dots (order of derivative according to color) and the original data as square points. The triangles in Panel (a) show the predicted toroid resonances from Ref. \cite{Cao7aPRC}.}
\label{Fig4}
\end{figure*}

After testing the AI method with model results, we apply it to the experimental data. The results of this process are shown in Fig \ref{Fig4} and follow the same nomenclature of Fig. \ref{Fig3} (panel c). The results of Cao \cite{Cao7aPRC} (panel a) are compared to the results of Hannaman \cite{Hannaman7aPRC} (panel b). For the former we obtain six peaks which approximately reproduce the predicted $114$ MeV and $138$ MeV toroid resonances (black triangles) in the region of high statistics. The peak at $126$ MeV is not reproduced independently, but belongs to the wide fourth Gaussian. The situation for the Hannaman yield data is similar with also six peaks. Interestingly, the resonances are close to the aforementioned found by Cao and collaborators. Our results indicate important underlying structures in the experimental datasets.\par
\begin{figure}[ht]                                        
\includegraphics[width=8.0 cm]{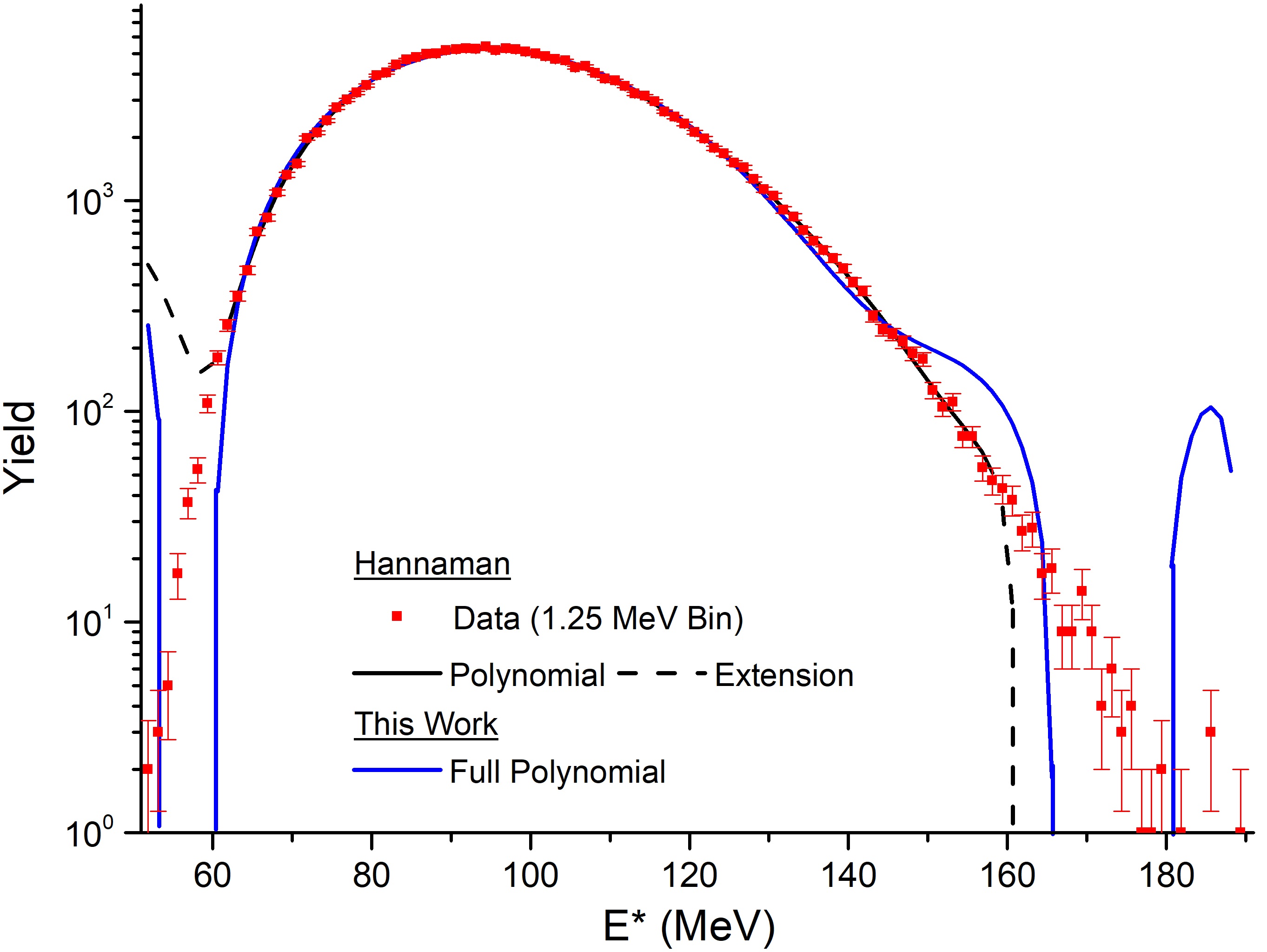}
\caption{(Color online) The polynomial fit of Ref. \cite{Hannaman7aPRC} (full black line) and its extension to the full data range (dashed black line). Our polynomial fit of the full data (blue line) and the original data as square points.}
\label{Fig10}
\end{figure}

This is illustrated in Fig. \ref{Fig10}, where the polynomial of Ref. \cite{Hannaman7aPRC} (full black line) is plotted and extended to cover the whole range of the data (dashed black line). An additional polynomial fit whose parameters are chosen to fit the full data range is also shown (blue line).  We can clearly observe that the polynomial is substantially different at the lower yield extremes of the data and creates two peaks, one positive and one negative, while our polynomial yields three peaks. Furthermore, there are minor inflections of the polynomial curves, which fluctuate around the data, but do not reflect real resonant structure. \par
\begin{figure*}[ht]                                        
\includegraphics[width=12cm]{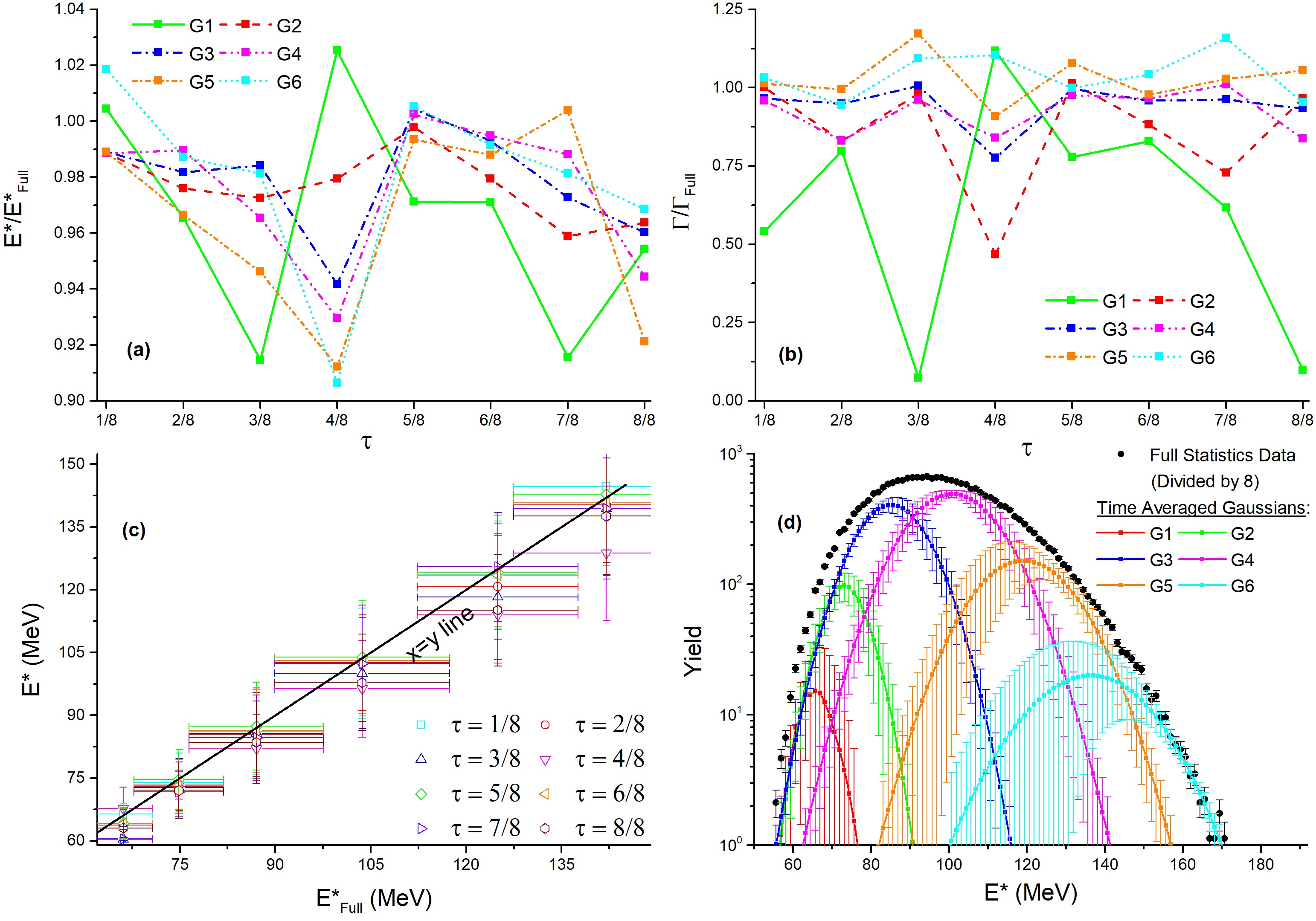}
\caption{(Color online)  AI calculations of the time-ordered partitions from the Hannaman data \cite{Hannaman7aPRC}. The excitation energies and widths divided by the full-statistics result for each Gaussian as functions of the time interval, in Panels (a) and (b). The excitation energies for each partition are also plotted against the full-statistics values in Panel (c), with the Half Widths at Half Maximum as error bars. The average Gaussians from all partitions, along with their standard deviations are shown in Panel (d).}
\label{Fig8}
\end{figure*}

\begin{figure*}[ht]                                        
\includegraphics[width=11.7 cm]{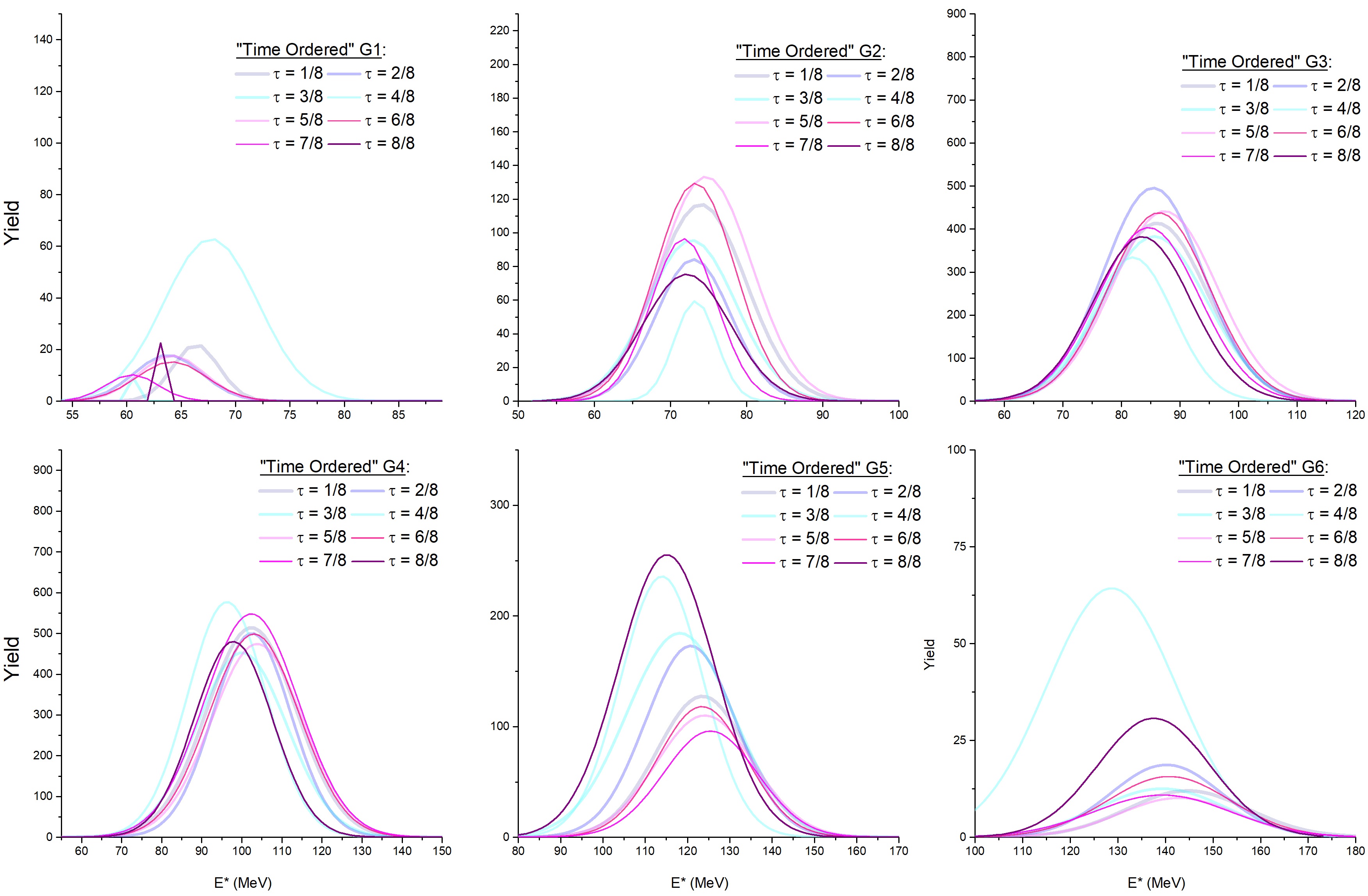}
\caption{(Color online) Evolution of AI Gaussians along the time-ordered partitions from the Hannaman data \cite{Hannaman7aPRC}. Each panel shows the evolution of one Gaussian, with different colors and fades, according to the key.}
\label{Fig9}
\end{figure*}

A point of contention between Ref.s \cite{Hannaman7aPRC,HannamanEPJ} and \cite{Cao7aPRC} is the statistical significance of the results. Specifically, Hannaman and collaborators claim that peaks in the data correspond to statistical fluctuations and thus, represent the background and not real toroidal states. To demonstrate the statistical significance of our results, we use the 186097 un-binned excitation energy points of Ref. \cite{Hannaman7aPRC}, which we divide into eight groups with equal size. The division takes place according to the time that were experimentally obtained, i.e., the $\tau=1/8$ partition is taken earlier than the $\tau=2/8$ and so on.  We then,  construct excitation energy spectra by binning the points of each time-ordered partition with a 1.25 MeV bin and apply our AI method to them. Our results are shown in Fig.s \ref{Fig8} and \ref{Fig9}.\par
In Fig. \ref{Fig8}, we show the evolution of the excitation energies and widths scaled by the full statistics results of each Gaussians along the time-ordered partitions in panels (a) and (b), respectively. We note that the excitation energies of each partition fluctuate around the full statistics value, with a only variation of $\sim 2.6 \%$. Their common tendencies might correspond to minor experimental shifts, which in our method are mostly compensated by the larger fluctuation of the widths. In panel (c), the excitation energies of the Gaussians for each partition are plotted one by one against the excitation energies of the full statistics data (Fig. \ref{Fig4}, Panel b). The error bars correspond to the Gaussian Half Widths at Half Maximum (HWHM) of the peaks.\par
The average Gaussian curves and standard deviations from all the partitions are presented in Fig. \ref{Fig8} Panel (d), while the evolution of each peak is shown in Fig. \ref{Fig9} (panels a-f) with different colors and fadings. The similarity between Fig. \ref{Fig4} Panel (b) and Fig. \ref{Fig8} Panel (d), the proximity of the points in \ref{Fig8} Panel (c) to the $x=y$ line and the relatively little change of each Gaussian in all the partitions in Fig. \ref{Fig9}, illustrate rather clearly the statistical significance of our results. Our method does not seem to depend on small statistical fluctuations of the data. The gross shape of all of the time-ordered subsets is similar, while small fluctuations are within their error bar. These differences will be very prominent when a polynomial is subtracted from the data, as done in Ref. \cite{Hannaman7aPRC}. Nevertheless, some fluctuations can be observed at regions of low statistics, i.e., at the tails of the distributions. We then demonstrate that, with this method, we can recover real statistically significant structure of the data, provided good initial conditions, proper choice of number of Gaussians and enough statistics.
\begin{figure*}[ht]                                        
\includegraphics[width=11.5 cm]{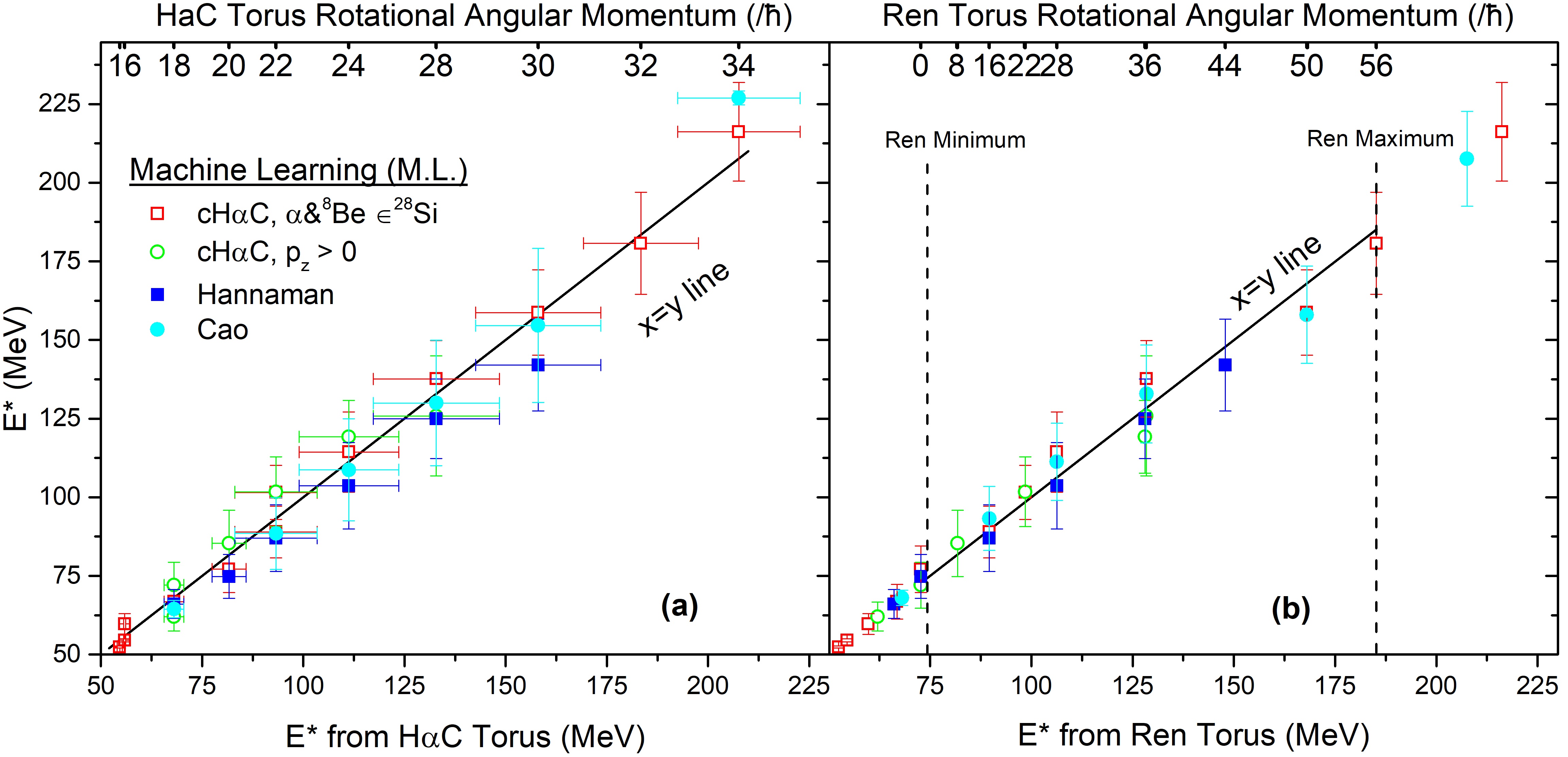}
\caption{(Color online) Collective machine learning excitation energy results for the theoretical and experimental datasets \cite{Cao7aPRC,Hannaman7aPRC} plotted against the results of the H$\alpha$C torus \cite{ZhengHac2021} test data (panel a) and the resonances reported in Ref. \cite{Ren2020} (panel b). The Half Widths at Half Maximum (HWHM's) as obtained from the machine learning Gaussians, are used as error bars for each peak. The rotational angular momenta corresponding to the horizontal axis excitation energies are noted on the top of each panel.}
\label{Fig6}
\end{figure*}
The machine learning results are collectively summarized in Fig.s \ref{Fig6} and \ref{Fig7}. In Fig. \ref{Fig6}, we plot the centroid values of the Gaussian peaks for each dataset (open symbols for theory and full for experiment) against the results for the rotating silicon training data (panel a) and the resonances reported in Ref. \cite{Ren2020}, along with their corresponding angular momenta on the top. We note that the $0\hbar$ state at $\sim 72$ MeV is the lowest toroidal state predicted in Ref. \cite{Ren2020}, where the distortion and binding energies of the silicon are balanced. The Half Width at Half Maximum of each Gaussian is used as the error bar in each axis. The peaks are matched with the closest toroidal state energy, while for panel (b) the peaks outside the prediction limits (dashed lines) are positioned arbitrarily close to the $x=y$ line.\par 
We observe a similarity of the theoretical and experimental reaction data with the results for the rotating silicon. This striking feature is signified by the proximity of all the curves with the $x=y$ line and might suggest the toroid-like nature of the confirmed underlying structure. This approximate agreement for the collision systems with thermal and rotational motion, with the toroid purely rotational calculation originates from the quantization of the impact parameter and the acceptance of $7\alpha$ fragments. The collision results are also in agreement in the region $E^*\sim 65 \text{ (panel a), } 75 \text{ (panel b) } - 150$ MeV and approximately reproduce the expected resonances at $114$ MeV and $138$ MeV \cite{Cao7aPRC}. The resonances at $126$ MeV \cite{Cao7aPRC} and $143.18$ MeV \cite{Wong2014} are found between the resulting peaks.\par
Our observations of similarity between the results for all the available datasets, are also shown in Fig. \ref{Fig7}. There we plot the FWHM against the Excitation Energy centroids for each M.L. Gaussian, following the same process as in Fig. \ref{Fig8}. The arrows show areas of common excitation energy peaks in the region of $60-150$ MeV. All the results are within the width of the experimental resonances \cite{Cao7aPRC} and thus, we may conclude that all examined $^{28}$Si + $^{12}$C data posses a collective structure. The widths and lifetimes of the extracted peaks, are close to those of the Giant $E0$, $E1$ and $E2$ Resonances, taken from Ref. \cite{Youngbloood2009}.
\begin{figure}[ht]                                        
\includegraphics[width=8.5 cm]{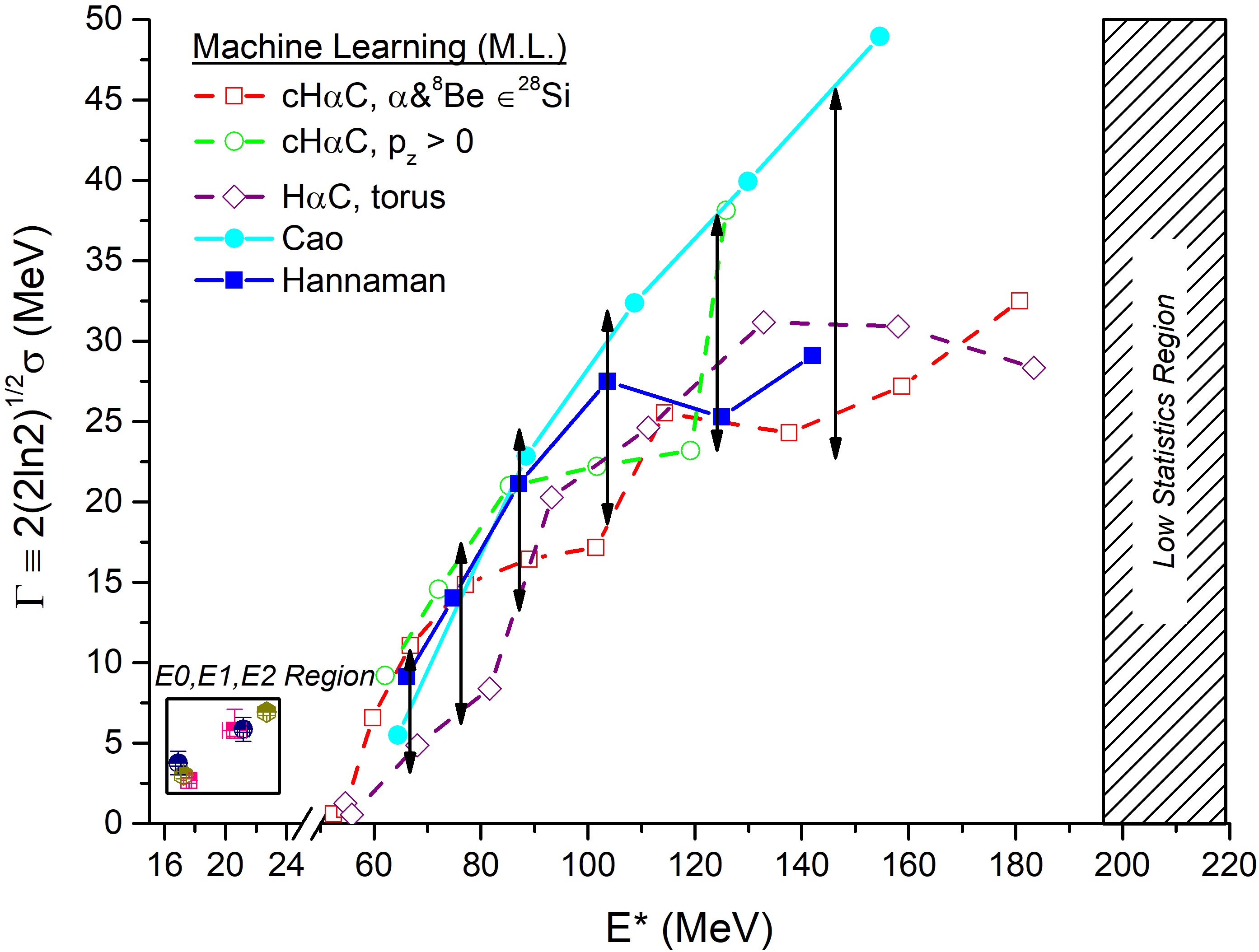}
\caption{(Color online) Collective machine learning Full Width at Half Maxima (FWHM, $\Gamma$) against the Excitation Energy ($<200$ MeV) centroids of each Gaussian for the theoretical \cite{ZhengHac2021} and experimental datasets \cite{Cao7aPRC,Hannaman7aPRC}, according to the key. For comparison, we also plot the values for the $E0$, $E1$ and $E2$ resonances of $^{28}$Si from Ref. \cite{Youngbloood2009}.}
\label{Fig7}
\end{figure}
\section{Conclusion}
\label{s5}
To summarize, we study the open problem of toroid-like resonances of $^{28}$Si$^*$ excited from the inverse kinematics reaction $^{28}$Si + $^{12}$C at $35$ MeV/A beam energy. We examine the experimental data of Ref. \cite{Cao7aPRC} and the theoretical data of Ref. \cite{ZhengHac2021}, which support the existence of toroid shapes, the experimental data of Ref.s \cite{Hannaman7aPRC,HannamanEPJ}, which disagree with the previous conclusions, as well as new theoretical H$\alpha$C model results for the collision.\par
The new theoretical results are obtained via the cH$\alpha$C model, which considers $\alpha$ particles as semi-classical fundamental degrees of freedom. We collect all the 7$\alpha$ events from free and $^8$Be-bound alpha particles. These are filtered based on two possible criteria, either by their origin in the projectile or by their momenta in the forward direction in the C.M. frame. The excitation spectrum is then calculated with Poisson standard errors.\par
To analyze the available data, we first try the polynomial fit previously used in the literature and note its drawbacks. We are then driven to develop a novel AI-based methodology. Our approach consists of transforming the cross section excitation spectrum into a discrete probability distribution and then employing the Unsupervised Machine Learning GMM technique to break down the distribution into a linear combination of Gaussian peaks. The method is tested with the H$\alpha$C rotating silicon data of Ref. \cite{ZhengHac2021}, with known Gaussian structure and then applied to the available reaction data. Our results approximately reproduce the predicted resonances and agree with the existence of underlying peaks. This is consistent with the conclusion drawn in Ref.s \cite{WadaCimento2025,NatowitzTalk}. Our study of the reaction with the H$\alpha$C model, suggests the presence of several interesting physical effects, ranging from collective motion, to thermodynamical and hydrodynamical phenomena.\\
\\ACKNOWLEDGMENTS\\
This work was supported in part by the United States Department of Energy under Grant $\#$DE-FG02-93ER40773.
\bibliographystyle{apsrev4-2}
\bibliography{references}{}

\providecommand{\noopsort}[1]{}\providecommand{\singleletter}[1]{#1}%
\begin{thebibliography}{36}%
\makeatletter
\providecommand \@ifxundefined [1]{%
 \@ifx{#1\undefined}
}%
\providecommand \@ifnum [1]{%
 \ifnum #1\expandafter \@firstoftwo
 \else \expandafter \@secondoftwo
 \fi
}%
\providecommand \@ifx [1]{%
 \ifx #1\expandafter \@firstoftwo
 \else \expandafter \@secondoftwo
 \fi
}%
\providecommand \natexlab [1]{#1}%
\providecommand \enquote  [1]{``#1''}%
\providecommand \bibnamefont  [1]{#1}%
\providecommand \bibfnamefont [1]{#1}%
\providecommand \citenamefont [1]{#1}%
\providecommand \href@noop [0]{\@secondoftwo}%
\providecommand \href [0]{\begingroup \@sanitize@url \@href}%
\providecommand \@href[1]{\@@startlink{#1}\@@href}%
\providecommand \@@href[1]{\endgroup#1\@@endlink}%
\providecommand \@sanitize@url [0]{\catcode `\\12\catcode `\$12\catcode `\&12\catcode `\#12\catcode `\^12\catcode `\_12\catcode `\%12\relax}%
\providecommand \@@startlink[1]{}%
\providecommand \@@endlink[0]{}%
\providecommand \url  [0]{\begingroup\@sanitize@url \@url }%
\providecommand \@url [1]{\endgroup\@href {#1}{\urlprefix }}%
\providecommand \urlprefix  [0]{URL }%
\providecommand \Eprint [0]{\href }%
\providecommand \doibase [0]{https://doi.org/}%
\providecommand \selectlanguage [0]{\@gobble}%
\providecommand \bibinfo  [0]{\@secondoftwo}%
\providecommand \bibfield  [0]{\@secondoftwo}%
\providecommand \translation [1]{[#1]}%
\providecommand \BibitemOpen [0]{}%
\providecommand \bibitemStop [0]{}%
\providecommand \bibitemNoStop [0]{.\EOS\space}%
\providecommand \EOS [0]{\spacefactor3000\relax}%
\providecommand \BibitemShut  [1]{\csname bibitem#1\endcsname}%
\let\auto@bib@innerbib\@empty
\bibitem [{\citenamefont {Basunia}\ and\ \citenamefont {Chakraborty}(2022)}]{NNDC2022}%
  \BibitemOpen
  \bibfield  {author} {\bibinfo {author} {\bibfnamefont {M.}~\bibnamefont {Basunia}}\ and\ \bibinfo {author} {\bibfnamefont {A.}~\bibnamefont {Chakraborty}},\ }\href@noop {} {\bibfield  {journal} {\bibinfo  {journal} {Nucl. Data Sheets}\ }\textbf {\bibinfo {volume} {186}} (\bibinfo {year} {2022})}\BibitemShut {NoStop}%
\bibitem [{\citenamefont {Ikeda}\ \emph {et~al.}(1968)\citenamefont {Ikeda}, \citenamefont {Takigawa},\ and\ \citenamefont {Horiuchi}}]{Ikeda1968}%
  \BibitemOpen
  \bibfield  {author} {\bibinfo {author} {\bibfnamefont {K.}~\bibnamefont {Ikeda}}, \bibinfo {author} {\bibfnamefont {N.}~\bibnamefont {Takigawa}},\ and\ \bibinfo {author} {\bibfnamefont {H.}~\bibnamefont {Horiuchi}},\ }\href@noop {} {\bibfield  {journal} {\bibinfo  {journal} {Progress of Theoretical Physics Supplement}\ }\textbf {\bibinfo {volume} {E68}},\ \bibinfo {pages} {464} (\bibinfo {year} {1968})}\BibitemShut {NoStop}%
\bibitem [{\citenamefont {Hoyle}(1954)}]{Hoyle1954}%
  \BibitemOpen
  \bibfield  {author} {\bibinfo {author} {\bibfnamefont {F.}~\bibnamefont {Hoyle}},\ }\href@noop {} {\bibfield  {journal} {\bibinfo  {journal} {Astrophys. J. Suppl.}\ }\textbf {\bibinfo {volume} {1}},\ \bibinfo {pages} {121} (\bibinfo {year} {1954})}\BibitemShut {NoStop}%
\bibitem [{\citenamefont {Borderie}\ \emph {et~al.}(2021)\citenamefont {Borderie}, \citenamefont {Raduta}, \citenamefont {De~Filippo}, \citenamefont {Geraci}, \citenamefont {Neindre}, \citenamefont {Cardella}, \citenamefont {Lanzalone}, \citenamefont {Lombardo}, \citenamefont {Lopez}, \citenamefont {Maiolino}, \citenamefont {Pagano}, \citenamefont {Papa}, \citenamefont {Pirrone}, \citenamefont {Rizzo},\ and\ \citenamefont {Russotto}}]{Borderie2021}%
  \BibitemOpen
  \bibfield  {author} {\bibinfo {author} {\bibfnamefont {B.}~\bibnamefont {Borderie}}, \bibinfo {author} {\bibfnamefont {A.}~\bibnamefont {Raduta}}, \bibinfo {author} {\bibfnamefont {E.}~\bibnamefont {De~Filippo}}, \bibinfo {author} {\bibfnamefont {E.}~\bibnamefont {Geraci}}, \bibinfo {author} {\bibfnamefont {N.~L.}\ \bibnamefont {Neindre}}, \bibinfo {author} {\bibfnamefont {G.}~\bibnamefont {Cardella}}, \bibinfo {author} {\bibfnamefont {G.}~\bibnamefont {Lanzalone}}, \bibinfo {author} {\bibfnamefont {I.}~\bibnamefont {Lombardo}}, \bibinfo {author} {\bibfnamefont {O.}~\bibnamefont {Lopez}}, \bibinfo {author} {\bibfnamefont {C.}~\bibnamefont {Maiolino}}, \bibinfo {author} {\bibfnamefont {A.}~\bibnamefont {Pagano}}, \bibinfo {author} {\bibfnamefont {M.}~\bibnamefont {Papa}}, \bibinfo {author} {\bibfnamefont {S.}~\bibnamefont {Pirrone}}, \bibinfo {author} {\bibfnamefont {F.}~\bibnamefont {Rizzo}},\ and\ \bibinfo {author} {\bibfnamefont {P.}~\bibnamefont {Russotto}},\ }\href@noop {} {\bibfield  {journal}
  {\bibinfo  {journal} {Symmetry}\ }\textbf {\bibinfo {volume} {13}} (\bibinfo {year} {2021})}\BibitemShut {NoStop}%
\bibitem [{\citenamefont {Brink}\ and\ \citenamefont {Castro}(1973)}]{Brink1973}%
  \BibitemOpen
  \bibfield  {author} {\bibinfo {author} {\bibfnamefont {D.}~\bibnamefont {Brink}}\ and\ \bibinfo {author} {\bibfnamefont {J.}~\bibnamefont {Castro}},\ }\href@noop {} {\bibfield  {journal} {\bibinfo  {journal} {Nuclear Physics A}\ }\textbf {\bibinfo {volume} {216}},\ \bibinfo {pages} {109} (\bibinfo {year} {1973})}\BibitemShut {NoStop}%
\bibitem [{\citenamefont {Zheng}\ and\ \citenamefont {Bonasera}(2011)}]{BonaseraAlphaMatter2011}%
  \BibitemOpen
  \bibfield  {author} {\bibinfo {author} {\bibfnamefont {H.}~\bibnamefont {Zheng}}\ and\ \bibinfo {author} {\bibfnamefont {A.}~\bibnamefont {Bonasera}},\ }\href@noop {} {\bibfield  {journal} {\bibinfo  {journal} {Phys. Rev. C}\ }\textbf {\bibinfo {volume} {83}},\ \bibinfo {pages} {057602} (\bibinfo {year} {2011})}\BibitemShut {NoStop}%
\bibitem [{\citenamefont {Bohr}(1951)}]{Bohr1951}%
  \BibitemOpen
  \bibfield  {author} {\bibinfo {author} {\bibfnamefont {A.}~\bibnamefont {Bohr}},\ }\href@noop {} {\bibfield  {journal} {\bibinfo  {journal} {Phys. Rev.}\ }\textbf {\bibinfo {volume} {81}},\ \bibinfo {pages} {331} (\bibinfo {year} {1951})}\BibitemShut {NoStop}%
\bibitem [{\citenamefont {Bohr}\ and\ \citenamefont {Mottelson}(1953)}]{BohrMottelson1953}%
  \BibitemOpen
  \bibfield  {author} {\bibinfo {author} {\bibfnamefont {A.}~\bibnamefont {Bohr}}\ and\ \bibinfo {author} {\bibfnamefont {B.~R.}\ \bibnamefont {Mottelson}},\ }\href@noop {} {\bibfield  {journal} {\bibinfo  {journal} {Phys. Rev.}\ }\textbf {\bibinfo {volume} {89}},\ \bibinfo {pages} {316} (\bibinfo {year} {1953})}\BibitemShut {NoStop}%
\bibitem [{\citenamefont {Griffin}\ and\ \citenamefont {Wheeler}(1957)}]{Wheeler1957}%
  \BibitemOpen
  \bibfield  {author} {\bibinfo {author} {\bibfnamefont {J.~J.}\ \bibnamefont {Griffin}}\ and\ \bibinfo {author} {\bibfnamefont {J.~A.}\ \bibnamefont {Wheeler}},\ }\href@noop {} {\bibfield  {journal} {\bibinfo  {journal} {Phys. Rev.}\ }\textbf {\bibinfo {volume} {108}},\ \bibinfo {pages} {311} (\bibinfo {year} {1957})}\BibitemShut {NoStop}%
\bibitem [{\citenamefont {Rainwater}(1950)}]{Rainwater1950}%
  \BibitemOpen
  \bibfield  {author} {\bibinfo {author} {\bibfnamefont {J.}~\bibnamefont {Rainwater}},\ }\href@noop {} {\bibfield  {journal} {\bibinfo  {journal} {Phys. Rev.}\ }\textbf {\bibinfo {volume} {79}},\ \bibinfo {pages} {432} (\bibinfo {year} {1950})}\BibitemShut {NoStop}%
\bibitem [{\citenamefont {Wong}(1972)}]{Wong1972}%
  \BibitemOpen
  \bibfield  {author} {\bibinfo {author} {\bibfnamefont {C.~Y.}\ \bibnamefont {Wong}},\ }\href@noop {} {\bibfield  {journal} {\bibinfo  {journal} {Physics Letters B}\ }\textbf {\bibinfo {volume} {41}},\ \bibinfo {pages} {446} (\bibinfo {year} {1972})}\BibitemShut {NoStop}%
\bibitem [{\citenamefont {Wong}(1973)}]{Wong1973}%
  \BibitemOpen
  \bibfield  {author} {\bibinfo {author} {\bibfnamefont {C.~Y.}\ \bibnamefont {Wong}},\ }\href@noop {} {\bibfield  {journal} {\bibinfo  {journal} {Annals of Physics}\ }\textbf {\bibinfo {volume} {77}},\ \bibinfo {pages} {279} (\bibinfo {year} {1973})}\BibitemShut {NoStop}%
\bibitem [{\citenamefont {Wong}(1978)}]{Wong1978}%
  \BibitemOpen
  \bibfield  {author} {\bibinfo {author} {\bibfnamefont {C.~Y.}\ \bibnamefont {Wong}},\ }\href@noop {} {\bibfield  {journal} {\bibinfo  {journal} {Phys. Rev. C}\ }\textbf {\bibinfo {volume} {17}},\ \bibinfo {pages} {331} (\bibinfo {year} {1978})}\BibitemShut {NoStop}%
\bibitem [{\citenamefont {Staszczak}\ and\ \citenamefont {Wong}(2014)}]{Wong2014}%
  \BibitemOpen
  \bibfield  {author} {\bibinfo {author} {\bibfnamefont {A.}~\bibnamefont {Staszczak}}\ and\ \bibinfo {author} {\bibfnamefont {C.~Y.}\ \bibnamefont {Wong}},\ }\href@noop {} {\bibfield  {journal} {\bibinfo  {journal} {Physics Letters B}\ }\textbf {\bibinfo {volume} {738}},\ \bibinfo {pages} {401} (\bibinfo {year} {2014})}\BibitemShut {NoStop}%
\bibitem [{\citenamefont {Cao}\ \emph {et~al.}(2019)\citenamefont {Cao}, \citenamefont {Kim}, \citenamefont {Schmidt}, \citenamefont {Hagel}, \citenamefont {Barbui}, \citenamefont {Gauthier}, \citenamefont {Wuenschel}, \citenamefont {Giuliani}, \citenamefont {Rodriguez}, \citenamefont {Kowalski}, \citenamefont {Zheng}, \citenamefont {Huang}, \citenamefont {Bonasera}, \citenamefont {Wada}, \citenamefont {Blando}, \citenamefont {Zhang}, \citenamefont {Wong}, \citenamefont {Staszczak}, \citenamefont {Ren}, \citenamefont {Wang}, \citenamefont {Zhang}, \citenamefont {Meng},\ and\ \citenamefont {Natowitz}}]{Cao7aPRC}%
  \BibitemOpen
  \bibfield  {author} {\bibinfo {author} {\bibfnamefont {X.~G.}\ \bibnamefont {Cao}}, \bibinfo {author} {\bibfnamefont {E.~J.}\ \bibnamefont {Kim}}, \bibinfo {author} {\bibfnamefont {K.}~\bibnamefont {Schmidt}}, \bibinfo {author} {\bibfnamefont {K.}~\bibnamefont {Hagel}}, \bibinfo {author} {\bibfnamefont {M.}~\bibnamefont {Barbui}}, \bibinfo {author} {\bibfnamefont {J.}~\bibnamefont {Gauthier}}, \bibinfo {author} {\bibfnamefont {S.}~\bibnamefont {Wuenschel}}, \bibinfo {author} {\bibfnamefont {G.}~\bibnamefont {Giuliani}}, \bibinfo {author} {\bibfnamefont {M.~R.~D.}\ \bibnamefont {Rodriguez}}, \bibinfo {author} {\bibfnamefont {S.}~\bibnamefont {Kowalski}}, \bibinfo {author} {\bibfnamefont {H.}~\bibnamefont {Zheng}}, \bibinfo {author} {\bibfnamefont {M.}~\bibnamefont {Huang}}, \bibinfo {author} {\bibfnamefont {A.}~\bibnamefont {Bonasera}}, \bibinfo {author} {\bibfnamefont {R.}~\bibnamefont {Wada}}, \bibinfo {author} {\bibfnamefont {N.}~\bibnamefont {Blando}}, \bibinfo {author} {\bibfnamefont {G.~Q.}\
  \bibnamefont {Zhang}}, \bibinfo {author} {\bibfnamefont {C.~Y.}\ \bibnamefont {Wong}}, \bibinfo {author} {\bibfnamefont {A.}~\bibnamefont {Staszczak}}, \bibinfo {author} {\bibfnamefont {Z.~X.}\ \bibnamefont {Ren}}, \bibinfo {author} {\bibfnamefont {Y.~K.}\ \bibnamefont {Wang}}, \bibinfo {author} {\bibfnamefont {S.~Q.}\ \bibnamefont {Zhang}}, \bibinfo {author} {\bibfnamefont {J.}~\bibnamefont {Meng}},\ and\ \bibinfo {author} {\bibfnamefont {J.~B.}\ \bibnamefont {Natowitz}},\ }\href@noop {} {\bibfield  {journal} {\bibinfo  {journal} {Phys. Rev. C}\ }\textbf {\bibinfo {volume} {99}},\ \bibinfo {pages} {014606} (\bibinfo {year} {2019})}\BibitemShut {NoStop}%
\bibitem [{\citenamefont {Ren}\ \emph {et~al.}(2020)\citenamefont {Ren}, \citenamefont {Zhao}, \citenamefont {Zhang},\ and\ \citenamefont {Meng}}]{Ren2020}%
  \BibitemOpen
  \bibfield  {author} {\bibinfo {author} {\bibfnamefont {Z.}~\bibnamefont {Ren}}, \bibinfo {author} {\bibfnamefont {P.}~\bibnamefont {Zhao}}, \bibinfo {author} {\bibfnamefont {S.}~\bibnamefont {Zhang}},\ and\ \bibinfo {author} {\bibfnamefont {J.}~\bibnamefont {Meng}},\ }\href@noop {} {\bibfield  {journal} {\bibinfo  {journal} {Nuclear Physics A}\ }\textbf {\bibinfo {volume} {996}},\ \bibinfo {pages} {121696} (\bibinfo {year} {2020})}\BibitemShut {NoStop}%
\bibitem [{\citenamefont {Hannaman}\ \emph {et~al.}(2024)\citenamefont {Hannaman}, \citenamefont {Harvey}, \citenamefont {McIntosh}, \citenamefont {Hagel}, \citenamefont {Abbott}, \citenamefont {Fentress}, \citenamefont {Gauthier}, \citenamefont {Hankins}, \citenamefont {Lui}, \citenamefont {McCann}, \citenamefont {McIntosh}, \citenamefont {Regener}, \citenamefont {Rider}, \citenamefont {Schultz}, \citenamefont {Sorensen}, \citenamefont {Tobar}, \citenamefont {Tobin},\ and\ \citenamefont {Yennello}}]{Hannaman7aPRC}%
  \BibitemOpen
  \bibfield  {author} {\bibinfo {author} {\bibfnamefont {A.}~\bibnamefont {Hannaman}}, \bibinfo {author} {\bibfnamefont {B.}~\bibnamefont {Harvey}}, \bibinfo {author} {\bibfnamefont {A.~B.}\ \bibnamefont {McIntosh}}, \bibinfo {author} {\bibfnamefont {K.}~\bibnamefont {Hagel}}, \bibinfo {author} {\bibfnamefont {A.}~\bibnamefont {Abbott}}, \bibinfo {author} {\bibfnamefont {A.}~\bibnamefont {Fentress}}, \bibinfo {author} {\bibfnamefont {J.}~\bibnamefont {Gauthier}}, \bibinfo {author} {\bibfnamefont {T.}~\bibnamefont {Hankins}}, \bibinfo {author} {\bibfnamefont {Y.-W.}\ \bibnamefont {Lui}}, \bibinfo {author} {\bibfnamefont {L.}~\bibnamefont {McCann}}, \bibinfo {author} {\bibfnamefont {L.~A.}\ \bibnamefont {McIntosh}}, \bibinfo {author} {\bibfnamefont {S.}~\bibnamefont {Regener}}, \bibinfo {author} {\bibfnamefont {R.}~\bibnamefont {Rider}}, \bibinfo {author} {\bibfnamefont {S.}~\bibnamefont {Schultz}}, \bibinfo {author} {\bibfnamefont {M.~Q.}\ \bibnamefont {Sorensen}}, \bibinfo {author} {\bibfnamefont
  {J.}~\bibnamefont {Tobar}}, \bibinfo {author} {\bibfnamefont {Z.~N.}\ \bibnamefont {Tobin}},\ and\ \bibinfo {author} {\bibfnamefont {S.~J.}\ \bibnamefont {Yennello}},\ }\href@noop {} {\bibfield  {journal} {\bibinfo  {journal} {Phys. Rev. C}\ }\textbf {\bibinfo {volume} {109}},\ \bibinfo {pages} {054615} (\bibinfo {year} {2024})}\BibitemShut {NoStop}%
\bibitem [{\citenamefont {{Hannaman, A.}}\ \emph {et~al.}(2024)\citenamefont {{Hannaman, A.}}, \citenamefont {{Harvey, B.}}, \citenamefont {{McIntosh, A.B.}}, \citenamefont {{Hagel, K.}}, \citenamefont {{Bills, L.}}, \citenamefont {{Abbott, A.}}, \citenamefont {{Fentress, A.}}, \citenamefont {{Gauthier, J.}}, \citenamefont {{Hankins, T.}}, \citenamefont {{Lui, Y.-W.}}, \citenamefont {{Bills, L.}}, \citenamefont {{McIntosh, L.A.}}, \citenamefont {{Regener, S.}}, \citenamefont {{Rider, R.}}, \citenamefont {{Schultz, S.}}, \citenamefont {{Sorensen, M.Q.}}, \citenamefont {{Tobar, J.}}, \citenamefont {{Tobin, Z.}},\ and\ \citenamefont {{Yennello, S.}}}]{HannamanEPJ}%
  \BibitemOpen
  \bibfield  {author} {\bibinfo {author} {\bibnamefont {{Hannaman, A.}}}, \bibinfo {author} {\bibnamefont {{Harvey, B.}}}, \bibinfo {author} {\bibnamefont {{McIntosh, A.B.}}}, \bibinfo {author} {\bibnamefont {{Hagel, K.}}}, \bibinfo {author} {\bibnamefont {{Bills, L.}}}, \bibinfo {author} {\bibnamefont {{Abbott, A.}}}, \bibinfo {author} {\bibnamefont {{Fentress, A.}}}, \bibinfo {author} {\bibnamefont {{Gauthier, J.}}}, \bibinfo {author} {\bibnamefont {{Hankins, T.}}}, \bibinfo {author} {\bibnamefont {{Lui, Y.-W.}}}, \bibinfo {author} {\bibnamefont {{Bills, L.}}}, \bibinfo {author} {\bibnamefont {{McIntosh, L.A.}}}, \bibinfo {author} {\bibnamefont {{Regener, S.}}}, \bibinfo {author} {\bibnamefont {{Rider, R.}}}, \bibinfo {author} {\bibnamefont {{Schultz, S.}}}, \bibinfo {author} {\bibnamefont {{Sorensen, M.Q.}}}, \bibinfo {author} {\bibnamefont {{Tobar, J.}}}, \bibinfo {author} {\bibnamefont {{Tobin, Z.}}},\ and\ \bibinfo {author} {\bibnamefont {{Yennello, S.}}},\ }\href@noop {} {\bibfield  {journal} {\bibinfo
   {journal} {EPJ Web Conf.}\ }\textbf {\bibinfo {volume} {304}},\ \bibinfo {pages} {01001} (\bibinfo {year} {2024})}\BibitemShut {NoStop}%
\bibitem [{\citenamefont {Wada}(2025)}]{WadaCimento2025}%
  \BibitemOpen
  \bibfield  {author} {\bibinfo {author} {\bibfnamefont {R.}~\bibnamefont {Wada}},\ }\href@noop {} {\bibfield  {journal} {\bibinfo  {journal} {IL NUOVO CIMENTO 48 C}\ }\textbf {\bibinfo {volume} {48}} (\bibinfo {year} {2025})}\BibitemShut {NoStop}%
\bibitem [{\citenamefont {Natowitz}\ and\ \citenamefont {Wada}(2024)}]{NatowitzTalk}%
  \BibitemOpen
  \bibfield  {author} {\bibinfo {author} {\bibfnamefont {J.}~\bibnamefont {Natowitz}}\ and\ \bibinfo {author} {\bibfnamefont {R.}~\bibnamefont {Wada}},\ }in\ \href@noop {} {\emph {\bibinfo {booktitle} {International Workshop on Nuclear Dynamics in Heavy-Ion Reactions}}}\ (\bibinfo {address} {Zhuhai, China},\ \bibinfo {year} {2024})\ \bibinfo {note} {\url{https://docs.google.com/presentation/d/e/2PACX-1vSZOTyNl3woLdLc_AMIYKJahYGcyNUsu8JO_7bVd-07oOr7w_MPU-lJc1o7yCQRkwEGyTrcIKGieT8i/embed?slide=id.p1} [Last Accessed: (May 2025)]}\BibitemShut {NoStop}%
\bibitem [{\citenamefont {Zheng}\ and\ \citenamefont {Bonasera}(2021)}]{ZhengHac2021}%
  \BibitemOpen
  \bibfield  {author} {\bibinfo {author} {\bibfnamefont {H.}~\bibnamefont {Zheng}}\ and\ \bibinfo {author} {\bibfnamefont {A.}~\bibnamefont {Bonasera}},\ }\href@noop {} {\bibfield  {journal} {\bibinfo  {journal} {Symmetry}\ }\textbf {\bibinfo {volume} {13}},\ \bibinfo {pages} {1777} (\bibinfo {year} {2021})}\BibitemShut {NoStop}%
\bibitem [{\citenamefont {Depastas}\ \emph {et~al.}(2023)\citenamefont {Depastas}, \citenamefont {Sun}, \citenamefont {Zheng},\ and\ \citenamefont {Bonasera}}]{Depastas2023}%
  \BibitemOpen
  \bibfield  {author} {\bibinfo {author} {\bibfnamefont {T.}~\bibnamefont {Depastas}}, \bibinfo {author} {\bibfnamefont {S.~T.}\ \bibnamefont {Sun}}, \bibinfo {author} {\bibfnamefont {H.}~\bibnamefont {Zheng}},\ and\ \bibinfo {author} {\bibfnamefont {A.}~\bibnamefont {Bonasera}},\ }\href@noop {} {\bibfield  {journal} {\bibinfo  {journal} {Phys. Rev. C}\ }\textbf {\bibinfo {volume} {108}},\ \bibinfo {pages} {035806} (\bibinfo {year} {2023})}\BibitemShut {NoStop}%
\bibitem [{\citenamefont {{Depastas, Theodoros}}\ \emph {et~al.}(2024)\citenamefont {{Depastas, Theodoros}}, \citenamefont {{Sun, Shuting}}, \citenamefont {{He, Hongbin}}, \citenamefont {{Zheng, Hua}},\ and\ \citenamefont {{Bonasera, Aldo}}}]{Depastas2024EPJ}%
  \BibitemOpen
  \bibfield  {author} {\bibinfo {author} {\bibnamefont {{Depastas, Theodoros}}}, \bibinfo {author} {\bibnamefont {{Sun, Shuting}}}, \bibinfo {author} {\bibnamefont {{He, Hongbin}}}, \bibinfo {author} {\bibnamefont {{Zheng, Hua}}},\ and\ \bibinfo {author} {\bibnamefont {{Bonasera, Aldo}}},\ }\href@noop {} {\bibfield  {journal} {\bibinfo  {journal} {EPJ Web Conf.}\ }\textbf {\bibinfo {volume} {304}},\ \bibinfo {pages} {02004} (\bibinfo {year} {2024})}\BibitemShut {NoStop}%
\bibitem [{\citenamefont {Depastas}\ \emph {et~al.}(2025)\citenamefont {Depastas}, \citenamefont {Sun}, \citenamefont {He}, \citenamefont {Zheng},\ and\ \citenamefont {Bonasera}}]{Depastas2024Plb}%
  \BibitemOpen
  \bibfield  {author} {\bibinfo {author} {\bibfnamefont {T.}~\bibnamefont {Depastas}}, \bibinfo {author} {\bibfnamefont {S.}~\bibnamefont {Sun}}, \bibinfo {author} {\bibfnamefont {H.}~\bibnamefont {He}}, \bibinfo {author} {\bibfnamefont {H.}~\bibnamefont {Zheng}},\ and\ \bibinfo {author} {\bibfnamefont {A.}~\bibnamefont {Bonasera}},\ }\href@noop {} {\bibfield  {journal} {\bibinfo  {journal} {Physics Letters B}\ }\textbf {\bibinfo {volume} {860}},\ \bibinfo {pages} {139180} (\bibinfo {year} {2025})}\BibitemShut {NoStop}%
\bibitem [{\citenamefont {Bass}(1977)}]{Bass1977}%
  \BibitemOpen
  \bibfield  {author} {\bibinfo {author} {\bibfnamefont {R.}~\bibnamefont {Bass}},\ }\href@noop {} {\bibfield  {journal} {\bibinfo  {journal} {Phys. Rev. Lett.}\ }\textbf {\bibinfo {volume} {39}},\ \bibinfo {pages} {265} (\bibinfo {year} {1977})}\BibitemShut {NoStop}%
\bibitem [{\citenamefont {Sanctis}\ \emph {et~al.}(2008)\citenamefont {Sanctis}, \citenamefont {Masotti}, \citenamefont {Bruno}, \citenamefont {D'Agostino}, \citenamefont {Geraci}, \citenamefont {Vannini},\ and\ \citenamefont {Bonasera}}]{DeSanctis2009}%
  \BibitemOpen
  \bibfield  {author} {\bibinfo {author} {\bibfnamefont {J.~D.}\ \bibnamefont {Sanctis}}, \bibinfo {author} {\bibfnamefont {M.}~\bibnamefont {Masotti}}, \bibinfo {author} {\bibfnamefont {M.}~\bibnamefont {Bruno}}, \bibinfo {author} {\bibfnamefont {M.}~\bibnamefont {D'Agostino}}, \bibinfo {author} {\bibfnamefont {E.}~\bibnamefont {Geraci}}, \bibinfo {author} {\bibfnamefont {G.}~\bibnamefont {Vannini}},\ and\ \bibinfo {author} {\bibfnamefont {A.}~\bibnamefont {Bonasera}},\ }\href@noop {} {\bibfield  {journal} {\bibinfo  {journal} {Journal of Physics G: Nuclear and Particle Physics}\ }\textbf {\bibinfo {volume} {36}},\ \bibinfo {pages} {015101} (\bibinfo {year} {2008})}\BibitemShut {NoStop}%
\bibitem [{\citenamefont {Hopfield}(1982)}]{Hopfiled1982Nobel2024}%
  \BibitemOpen
  \bibfield  {author} {\bibinfo {author} {\bibfnamefont {J.~J.}\ \bibnamefont {Hopfield}},\ }\href@noop {} {\bibfield  {journal} {\bibinfo  {journal} {Proceedings of the National Academy of Sciences}\ }\textbf {\bibinfo {volume} {79}},\ \bibinfo {pages} {2554} (\bibinfo {year} {1982})}\BibitemShut {NoStop}%
\bibitem [{\citenamefont {Ackley}\ \emph {et~al.}(1985)\citenamefont {Ackley}, \citenamefont {Hinton},\ and\ \citenamefont {Sejnowski}}]{Ackley1985Nobel2024}%
  \BibitemOpen
  \bibfield  {author} {\bibinfo {author} {\bibfnamefont {D.~H.}\ \bibnamefont {Ackley}}, \bibinfo {author} {\bibfnamefont {G.~E.}\ \bibnamefont {Hinton}},\ and\ \bibinfo {author} {\bibfnamefont {T.~J.}\ \bibnamefont {Sejnowski}},\ }\href@noop {} {\bibfield  {journal} {\bibinfo  {journal} {Cognitive Science}\ }\textbf {\bibinfo {volume} {9}},\ \bibinfo {pages} {147} (\bibinfo {year} {1985})}\BibitemShut {NoStop}%
\bibitem [{\citenamefont {Jumper}\ \emph {et~al.}(2021)\citenamefont {Jumper}, \citenamefont {Evans}, \citenamefont {Pritzel}, \citenamefont {Green}, \citenamefont {Figurnov}, \citenamefont {Ronneberger}, \citenamefont {Tunyasuvunakool}, \citenamefont {Bates}, \citenamefont {Žídek}, \citenamefont {Potapenko}, \citenamefont {Bridgland}, \citenamefont {Meyer}, \citenamefont {Kohl}, \citenamefont {Ballard}, \citenamefont {Cowie}, \citenamefont {Romera-Paredes}, \citenamefont {Nikolov}, \citenamefont {Jain}, \citenamefont {Adler}, \citenamefont {Back}, \citenamefont {Petersen}, \citenamefont {Reiman}, \citenamefont {Clancy}, \citenamefont {Zielinski}, \citenamefont {Steinegger}, \citenamefont {Pacholska}, \citenamefont {Berghammer}, \citenamefont {Bodenstein}, \citenamefont {Silver}, \citenamefont {Vinyals}, \citenamefont {Senior}, \citenamefont {K.~Kavukcuoglu},\ and\ \citenamefont {Hassabis}}]{Hassabis2021Nobel2024}%
  \BibitemOpen
  \bibfield  {author} {\bibinfo {author} {\bibfnamefont {J.}~\bibnamefont {Jumper}}, \bibinfo {author} {\bibfnamefont {R.}~\bibnamefont {Evans}}, \bibinfo {author} {\bibfnamefont {A.}~\bibnamefont {Pritzel}}, \bibinfo {author} {\bibfnamefont {T.}~\bibnamefont {Green}}, \bibinfo {author} {\bibfnamefont {M.}~\bibnamefont {Figurnov}}, \bibinfo {author} {\bibfnamefont {O.}~\bibnamefont {Ronneberger}}, \bibinfo {author} {\bibfnamefont {K.}~\bibnamefont {Tunyasuvunakool}}, \bibinfo {author} {\bibfnamefont {R.}~\bibnamefont {Bates}}, \bibinfo {author} {\bibfnamefont {A.}~\bibnamefont {Žídek}}, \bibinfo {author} {\bibfnamefont {A.}~\bibnamefont {Potapenko}}, \bibinfo {author} {\bibfnamefont {A.}~\bibnamefont {Bridgland}}, \bibinfo {author} {\bibfnamefont {C.}~\bibnamefont {Meyer}}, \bibinfo {author} {\bibfnamefont {S.}~\bibnamefont {Kohl}}, \bibinfo {author} {\bibfnamefont {A.}~\bibnamefont {Ballard}}, \bibinfo {author} {\bibfnamefont {A.}~\bibnamefont {Cowie}}, \bibinfo {author} {\bibfnamefont {B.}~\bibnamefont
  {Romera-Paredes}}, \bibinfo {author} {\bibfnamefont {S.}~\bibnamefont {Nikolov}}, \bibinfo {author} {\bibfnamefont {R.}~\bibnamefont {Jain}}, \bibinfo {author} {\bibfnamefont {J.}~\bibnamefont {Adler}}, \bibinfo {author} {\bibfnamefont {T.}~\bibnamefont {Back}}, \bibinfo {author} {\bibfnamefont {S.}~\bibnamefont {Petersen}}, \bibinfo {author} {\bibfnamefont {D.}~\bibnamefont {Reiman}}, \bibinfo {author} {\bibfnamefont {E.}~\bibnamefont {Clancy}}, \bibinfo {author} {\bibfnamefont {M.}~\bibnamefont {Zielinski}}, \bibinfo {author} {\bibfnamefont {M.}~\bibnamefont {Steinegger}}, \bibinfo {author} {\bibfnamefont {M.}~\bibnamefont {Pacholska}}, \bibinfo {author} {\bibfnamefont {T.}~\bibnamefont {Berghammer}}, \bibinfo {author} {\bibfnamefont {S.}~\bibnamefont {Bodenstein}}, \bibinfo {author} {\bibfnamefont {D.}~\bibnamefont {Silver}}, \bibinfo {author} {\bibfnamefont {O.}~\bibnamefont {Vinyals}}, \bibinfo {author} {\bibfnamefont {A.}~\bibnamefont {Senior}}, \bibinfo {author} {\bibfnamefont {P.~K.}\ \bibnamefont
  {K.~Kavukcuoglu}},\ and\ \bibinfo {author} {\bibfnamefont {D.}~\bibnamefont {Hassabis}},\ }\href@noop {} {\bibfield  {journal} {\bibinfo  {journal} {Nature}\ }\textbf {\bibinfo {volume} {596}},\ \bibinfo {pages} {583–589} (\bibinfo {year} {2021})}\BibitemShut {NoStop}%
\bibitem [{\citenamefont {Jothilakshmi}\ and\ \citenamefont {Gudivada}(2016)}]{GMMBOOK}%
  \BibitemOpen
  \bibfield  {author} {\bibinfo {author} {\bibfnamefont {S.}~\bibnamefont {Jothilakshmi}}\ and\ \bibinfo {author} {\bibfnamefont {V.}~\bibnamefont {Gudivada}},\ }in\ \href@noop {} {\emph {\bibinfo {booktitle} {Cognitive Computing: Theory and Applications}}},\ \bibinfo {series} {Handbook of Statistics}, Vol.~\bibinfo {volume} {35},\ \bibinfo {editor} {edited by\ \bibinfo {editor} {\bibfnamefont {V.~N.}\ \bibnamefont {Gudivada}}, \bibinfo {editor} {\bibfnamefont {V.~V.}\ \bibnamefont {Raghavan}}, \bibinfo {editor} {\bibfnamefont {V.}~\bibnamefont {Govindaraju}},\ and\ \bibinfo {editor} {\bibfnamefont {C.}~\bibnamefont {Rao}}}\ (\bibinfo  {publisher} {Elsevier},\ \bibinfo {year} {2016})\ pp.\ \bibinfo {pages} {301--340}\BibitemShut {NoStop}%
\bibitem [{\citenamefont {Pedregosa}\ \emph {et~al.}(2011)\citenamefont {Pedregosa}, \citenamefont {Varoquaux}, \citenamefont {Gramfort}, \citenamefont {Michel}, \citenamefont {Thirion}, \citenamefont {Grisel}, \citenamefont {Blondel}, \citenamefont {Prettenhofer}, \citenamefont {Weiss}, \citenamefont {Dubourg}, \citenamefont {Vanderplas}, \citenamefont {Passos}, \citenamefont {Cournapeau}, \citenamefont {Brucher}, \citenamefont {Perrot},\ and\ \citenamefont {Duchesnay}}]{scikit-learn}%
  \BibitemOpen
  \bibfield  {author} {\bibinfo {author} {\bibfnamefont {F.}~\bibnamefont {Pedregosa}}, \bibinfo {author} {\bibfnamefont {G.}~\bibnamefont {Varoquaux}}, \bibinfo {author} {\bibfnamefont {A.}~\bibnamefont {Gramfort}}, \bibinfo {author} {\bibfnamefont {V.}~\bibnamefont {Michel}}, \bibinfo {author} {\bibfnamefont {B.}~\bibnamefont {Thirion}}, \bibinfo {author} {\bibfnamefont {O.}~\bibnamefont {Grisel}}, \bibinfo {author} {\bibfnamefont {M.}~\bibnamefont {Blondel}}, \bibinfo {author} {\bibfnamefont {P.}~\bibnamefont {Prettenhofer}}, \bibinfo {author} {\bibfnamefont {R.}~\bibnamefont {Weiss}}, \bibinfo {author} {\bibfnamefont {V.}~\bibnamefont {Dubourg}}, \bibinfo {author} {\bibfnamefont {J.}~\bibnamefont {Vanderplas}}, \bibinfo {author} {\bibfnamefont {A.}~\bibnamefont {Passos}}, \bibinfo {author} {\bibfnamefont {D.}~\bibnamefont {Cournapeau}}, \bibinfo {author} {\bibfnamefont {M.}~\bibnamefont {Brucher}}, \bibinfo {author} {\bibfnamefont {M.}~\bibnamefont {Perrot}},\ and\ \bibinfo {author} {\bibfnamefont
  {E.}~\bibnamefont {Duchesnay}},\ }\href@noop {} {\bibfield  {journal} {\bibinfo  {journal} {Journal of Machine Learning Research}\ }\textbf {\bibinfo {volume} {12}},\ \bibinfo {pages} {2825} (\bibinfo {year} {2011})}\BibitemShut {NoStop}%
\bibitem [{\citenamefont {Arthur}\ and\ \citenamefont {Vassilvitskii}(2007)}]{kmeans}%
  \BibitemOpen
  \bibfield  {author} {\bibinfo {author} {\bibfnamefont {D.}~\bibnamefont {Arthur}}\ and\ \bibinfo {author} {\bibfnamefont {S.}~\bibnamefont {Vassilvitskii}},\ }in\ \href@noop {} {\emph {\bibinfo {booktitle} {Proceedings of the Eighteenth Annual ACM-SIAM Symposium on Discrete Algorithms}}},\ \bibinfo {series and number} {SODA '07}\ (\bibinfo  {publisher} {Society for Industrial and Applied Mathematics},\ \bibinfo {address} {USA},\ \bibinfo {year} {2007})\ p.\ \bibinfo {pages} {1027–1035}\BibitemShut {NoStop}%
\bibitem [{\citenamefont {Andrae}\ \emph {et~al.}(2010)\citenamefont {Andrae}, \citenamefont {Schulze-Hartung},\ and\ \citenamefont {Melchior}}]{chi2Ref}%
  \BibitemOpen
  \bibfield  {author} {\bibinfo {author} {\bibfnamefont {R.}~\bibnamefont {Andrae}}, \bibinfo {author} {\bibfnamefont {T.}~\bibnamefont {Schulze-Hartung}},\ and\ \bibinfo {author} {\bibfnamefont {P.}~\bibnamefont {Melchior}},\ }\href@noop {} {} (\bibinfo {year} {2010}),\ \Eprint {https://arxiv.org/abs/1012.3754} {arXiv:1012.3754 [astro-ph.IM]} \BibitemShut {NoStop}%
\bibitem [{\citenamefont {Papa}\ \emph {et~al.}(2001)\citenamefont {Papa}, \citenamefont {Maruyama},\ and\ \citenamefont {Bonasera}}]{BonaseraCoMD}%
  \BibitemOpen
  \bibfield  {author} {\bibinfo {author} {\bibfnamefont {M.}~\bibnamefont {Papa}}, \bibinfo {author} {\bibfnamefont {T.}~\bibnamefont {Maruyama}},\ and\ \bibinfo {author} {\bibfnamefont {A.}~\bibnamefont {Bonasera}},\ }\href@noop {} {\bibfield  {journal} {\bibinfo  {journal} {Phys. Rev. C}\ }\textbf {\bibinfo {volume} {64}},\ \bibinfo {pages} {024612} (\bibinfo {year} {2001})}\BibitemShut {NoStop}%
\bibitem [{\citenamefont {Bonasera}\ \emph {et~al.}(1990)\citenamefont {Bonasera}, \citenamefont {Colonna}, \citenamefont {{Di Toro}}, \citenamefont {Gulminelli},\ and\ \citenamefont {Wolter}}]{BonaseraDissipative1990}%
  \BibitemOpen
  \bibfield  {author} {\bibinfo {author} {\bibfnamefont {A.}~\bibnamefont {Bonasera}}, \bibinfo {author} {\bibfnamefont {M.}~\bibnamefont {Colonna}}, \bibinfo {author} {\bibfnamefont {M.}~\bibnamefont {{Di Toro}}}, \bibinfo {author} {\bibfnamefont {F.}~\bibnamefont {Gulminelli}},\ and\ \bibinfo {author} {\bibfnamefont {H.}~\bibnamefont {Wolter}},\ }\href@noop {} {\bibfield  {journal} {\bibinfo  {journal} {Physics Letters B}\ }\textbf {\bibinfo {volume} {244}},\ \bibinfo {pages} {169} (\bibinfo {year} {1990})}\BibitemShut {NoStop}%
\bibitem [{\citenamefont {Chen}\ \emph {et~al.}(2009)\citenamefont {Chen}, \citenamefont {Lui}, \citenamefont {Clark}, \citenamefont {Tokimoto},\ and\ \citenamefont {Youngblood}}]{Youngbloood2009}%
  \BibitemOpen
  \bibfield  {author} {\bibinfo {author} {\bibfnamefont {X.}~\bibnamefont {Chen}}, \bibinfo {author} {\bibfnamefont {Y.~W.}\ \bibnamefont {Lui}}, \bibinfo {author} {\bibfnamefont {H.~L.}\ \bibnamefont {Clark}}, \bibinfo {author} {\bibfnamefont {Y.}~\bibnamefont {Tokimoto}},\ and\ \bibinfo {author} {\bibfnamefont {D.~H.}\ \bibnamefont {Youngblood}},\ }\href@noop {} {\bibfield  {journal} {\bibinfo  {journal} {Phys. Rev. C}\ }\textbf {\bibinfo {volume} {80}},\ \bibinfo {pages} {014312} (\bibinfo {year} {2009})}\BibitemShut {NoStop}%
\end{thebibliography}%
\end{document}